\documentclass[times,tighten,twocolumn]{aastex631}

\usepackage{booktabs}
\usepackage{xspace}
\graphicspath{{./}{figures/}}

\newcommand{\teff}{$T_{\mathrm{eff}}$\xspace}
\newcommand{\logg}{$\log(g)$\xspace}
\newcommand{\fourA}{$4\langle \mathrm{A}\rangle$\xspace}

\received{August 21, 2021}
\revised{January 10, 2022}
\accepted{January 11, 2022}

\submitjournal{ApJ}

\shorttitle{New DBVs from the Sloan Digital Sky Survey}
\shortauthors{Vanderbosch et al.}

\begin{document}

\title{The Pulsating Helium-Atmosphere White Dwarfs I: New DBVs from the Sloan Digital Sky Survey}


\correspondingauthor{Zach Vanderbosch}
\email{zvanderb@caltech.edu}

\author[0000-0002-0853-3464]{Zachary P. Vanderbosch}
\affil{Division of Physics, Mathematics, and Astronomy, California Institute of Technology, Pasadena, CA 91125, USA}

\author[0000-0001-5941-2286]{J.~J. Hermes}
\affil{Department of Astronomy, Boston University, Boston, MA-02215, USA}

\author[0000-0003-0181-2521]{Don E. Winget}
\affil{Department of Astronomy, University of Texas at Austin, Austin, TX-78712, USA}
\affil{McDonald Observatory, Fort Davis, TX-79734, USA}

\author[0000-0002-6748-1748]{Michael H. Montgomery}
\affil{Department of Astronomy, University of Texas at Austin, Austin, TX-78712, USA}
\affil{McDonald Observatory, Fort Davis, TX-79734, USA}

\author[0000-0002-0656-032X]{Keaton J. Bell}
\affil{DIRAC Institute, Department of Astronomy, University of Washington, Seattle, WA-98195, USA}
\affil{NSF Astronomy and Astrophysics Postdoctoral Fellow}

\author{Atsuko Nitta}
\affil{Gemini Observatory, 670 North A’ohoku Place, Hilo, HI 96720, USA}

\author[0000-0002-7470-5703]{S.~O. Kepler}
\affil{Instituto de F\'{i}sica, Universidade Federal do Rio Grande do Sul, 91501-900 Porto-Alegre, RS, Brazil}



\begin{abstract}
We present a dedicated search for new pulsating helium-atmosphere (DBV) white dwarfs from the Sloan Digital Sky Survey using the McDonald 2.1m Otto Struve Telescope. In total we observed 55 DB and DBA white dwarfs with spectroscopic temperatures between 19,000 and 35,000\,K. We find 19 new DBVs and place upper limits on variability for the remaining 36 objects. In combination with previously known DBVs, we use these objects to provide an update to the empirical extent of the DB instability strip. With our sample of new DBVs, the red edge is better constrained, as we nearly double the number of DBVs known between 20,000 and 24,000\,K. We do not find any new DBVs hotter than PG\,0112+104, the current hottest DBV at $T_{\mathrm{eff}}\,{\approx}\,31,000$\,K, but do find pulsations in four DBVs with temperatures between 27,000 and 30,000\,K, improving empirical constraints on the poorly defined blue edge. We investigate the ensemble pulsation properties of all currently known DBVs, finding that the weighted mean period and total pulsation power exhibit trends with effective temperature that are qualitatively similar to the pulsating hydrogen-atmosphere white dwarfs.
\end{abstract}


\section{Introduction} \label{sec:intro}

White dwarf stars are the endpoint of stellar evolution for the vast majority of stars in our Galaxy. About 20\% of white dwarfs have optical spectra dominated by He lines, the DB and DBA stars \citep{Kleinman2013,Kepler_2019_1}. DBs show only He-I lines, while DBAs also show evidence for trace atmospheric H via H$\alpha$ or H$\beta$ absorption. As these white dwarfs cool monotonically throughout their lifetimes, they eventually reach the DB instability strip which covers effective temperatures, \teff, between 30,000 and 22,000\,K, at gravities of $\log(g\,[\mathrm{cm\,s^{-2]}}])=8.0$. Here, a He partial ionization region forms near the surface, and interior pulsations are driven to observable amplitudes. The exact location of the instability strip is mass dependent, with higher mass white dwarfs pulsating at higher temperatures. For the more common DA white dwarfs, those with H-dominated atmospheres, the instability strip occurs at lower temperatures ($12,500 > T_{\mathrm{eff}} > 10,500\,$K at $\log(g)=8.0$).

Following the detections of pulsations in DA white dwarfs \citep{Landolt1968}, pulsations in DBs were first predicted and then discovered by \citet{Winget1982} in GD\,358, the prototypical DBV. Since then, DBVs have remained difficult to find relative to their DAV counterparts due to faster cooling rates and the relative rarity of white dwarfs with He-dominated versus H-dominated atmospheres. Only 28 DBVs are currently known \citep[see][]{Corsico2019,Duan2021}. The largest boost in DBV numbers came from \citet{Nitta_2009_1}, who followed up candidates from the Sloan Digital Sky Survey (SDSS) using time series photometry from McDonald Observatory. Nitta et al. doubled the number from nine to 18, but most DBVs have been found just one or two at a time. Four pulsators have been identified as DBAVs so far due to detections of trace H \citep{Giammichele2018,Rolland2018}.

Increasing the number of known DBVs is beneficial for several reasons. More DBVs provides more candidates for which asteroseismic modeling can be used to probe the interior composition and structure of DBs (e.g., \citealt{Timmes2018,Charpinet2019}). These models place valuable boundary conditions on stellar evolution, particularly for DBs whose evolutionary origins are still debated. Below 30,000$-$40,000\,K, some DBs likely appear due to the convective dilution of a thin, primordial H layer ($M_{\mathrm{H}}=10^{-16}-10^{-14}\,M_{\odot}$) within the more massive He convection zone \citep{MacDonald1991,Rolland2018}, but this model cannot currently explain all DBs, especially those with large H contents whose H-layer ought to have survived convective dilution and remained as DAs. Other evolutionary channels, such as binary mergers \citep{Nather1981}, may account for some DBs, which would produce different internal structures that can be probed by asteroseismic models. To date, only a handful of DBVs have enough detected periods to perform asteroseismic modeling, such as GD\,358 \citep{Bischoff-Kim2019}, CBS\,114 \citep{Metcalfe2005}, PG\,0112+104 \citep{Hermes_2017_2}, KIC\,8626021 \citep{Bischoff-Kim2014,Giammichele2018}, TIC\,257459955 \citep{Bell2019}, and EPIC\,228782059 \citep{Duan2021}.

Finding more DBVs also increases the chance of finding candidates for stable pulsations. Long-term secular changes in pulsation periods can be used to measure white dwarf cooling rates \citep[e.g.,][]{Kepler2021}, while periodic variations in pulsation arrival times can be used to search for planetary companions \citep{Winget2003,Mullally2008}. Plasmon neutrino emission is thought to be the dominant source of energy loss in white dwarfs above 25,000\,K \citep{Winget2004}, so hot DBVs probe a unique parameter space not accessible by stable DAV pulsators. Fortunately, blue (hot) edge pulsating white dwarfs are expected to have more stable pulsations relative to red (cool) edge pulsators since the propagation cavities of their excited modes do not yet interact with the base of the convection zone \citep{Montgomery2020}. Therefore, from a theoretical perspective, stable DBVs are more likely to be found near the blue edge and probe relatively high neutrino emission rates. 

To date, only one DBV has been identified as a good candidate for stable pulsations and used to place preliminary constraints on its cooling rate \citep[PG\,1351+456,][]{Radaelli2011,Battich2016}, though it is likely too cool for neutrino emission to be the dominant source of energy loss. Of the two hottest DBVs, only PG\,0112+104 remains a good candidate for stable pulsations as EC\,20058$-$5234 has shown long-term period changes that cannot be accounted for by white dwarf cooling processes \citep{Delassio2013,Sullivan2017}. Unfortunately, even though PG\,0112+104 does exhibit a high degree of stability throughout its ${\approx}80$-day {\em K2} campaign with the {\em Kepler} spacecraft, the modes are very low amplitude ($<0.03$\% in the {\em Kepler} bandpass) and therefore difficult to detect and monitor with ground-based observations. Finding a hot DBV with stable pulsations that is both bright and relatively high amplitude would provide a unique opportunity to test white dwarf cooling theory.

Another use in finding more DBVs is placing empirical constraints on the extent of the DB instability strip. The location of the hot (blue) edge is strongly dependent on the convective efficiency at the base of the convection zone \citep{Fontaine2008,Corsico2009,vanGrootel_2017_1}, and theory suggests it might be used to calibrate the mixing length to pressure scale height ratio ($\alpha$) used in 1D mixing length theory \citep[e.g.,][]{Beauchamp1999}. The assumed convective efficiency has strong implications for the diffusion timescales of heavy metals accreted from disrupted planetary material, which are needed to accurately infer bulk planetary compositions \citep[e.g.,][]{Zuckerman2007,Dufour2010b,Melis2011,Farihi2013,Farihi_2016_2,Bauer2019, Xu_2019_2,Hoskin2020}. The cool (red) edge has traditionally defied a self-consistent theoretical description, but was most recently calculated by \citet{vanGrootel_2017_1} using an energy leakage criterion and found to be at 22,000\,K at $\log(g)=8.0$, about 1000\,K cooler than the observed red edge. 

Two main factors currently limit the ability to properly define the observed blue and red edges: (1) the small number of both hot and cool DBVs, and (2) the difficulty in obtaining accurate effective temperatures. The observed blue edge is currently defined by one object, PG\,0112+104 \citep{Shipman2002,Provencal2003,Hermes_2017_2} with $T_{\mathrm{eff}}\,{=}\,31{,}000\,$K and $\log(g)\,{=}\,7.8$ \citep{Dufour2010,Rolland2018}. PG\,0112+104 is about 2,000\,K hotter than the theoretical blue edge proposed by \citet{vanGrootel_2017_1}, suggesting a higher convective efficiency (higher $\alpha$) is required to explain the observed blue edge. Pulsations in PG\,0112+104 have extremely low amplitudes, with maximum amplitudes detected by {\em Kepler} observations of about $0.03\%$ \citep{Hermes_2017_2}. This suggests that many DBs that have historically been considered non-pulsators beyond the blue edge might indeed pulsate with amplitudes difficult to detect without extensive observations \citep[e.g.,][]{Castanheira2010}.

The atmospheric parameters for DBs (\teff, \logg, and H/He) are typically determined via a comparison of observed and model spectra \citep[the spectroscopic technique,][]{Eisenstein2006_2,Bergeron2011} or broadband photometry \citep[the photometric technique,][]{Bergeron1997,Genest2019_2}. Depending on which method and which set of atmospheric models are used to obtain atmospheric parameters, the observed red edge can vary in location between 20,000 and 23,000\,K, and is often defined by fewer than five objects. Similar effects on the DA instability strip have also been observed when comparing spectroscopic and photometric atmospheric parameters, with photometric parameters derived from Pan-STARRS1 {\em grizy} photometry being a few hundred degrees cooler on average within the DA instability strip compared to the spectroscopic technique \citep{Vincent2020, Bergeron2019}.  When defining the blue and red edges observationally, it is also important to find non-variable DBs near to or within the instability strip. \citet{Nitta_2009_1} reported several DBs not-observed-to-vary (NOV), but most had only one night of observations and variability limits above 0.5\%, leaving their NOV status uncertain. Many DBVs exhibit pulsation amplitudes below this limit, and destructive beating in multi-periodic pulsators can often make them appear non-variable in short single-night runs.

Even with more DBVs, however, the large uncertainties in their effective temperatures pose a significant challenge in defining the DB instability strip. DBV temperatures often vary by large amounts between studies due in large part to the insensitivity of He-I absorption lines to changes in \teff throughout the instability strip \citep[see Figure 2 of][]{Bergeron2011}. Peak He-I line strength occurs around 25,000\,K, which often leads to degenerate hot and cold solutions for DBVs, especially those with low signal-to-noise (S/N) observations. A common practice nowadays is to use broadband photometry, such as Sloan Digital Sky Survey (SDSS) $ugriz$ \citep[e.g.,][]{KK2015,Kepler_2019_1}, to try and break the degeneracy with a photometric fit, but this does not always guarantee the correct solution. Masses for DBs have also been found to be unreliable, though only at lower temperatures ($T_{\mathrm{eff}}\lesssim16,000\,$K), which is suspected to arise from an improper treatment of neutral broadening \citep[see][]{KK2015,Cukanovaite2018}. 

An additional source of temperature uncertainty comes from the presence of trace undetected H. This effect was first studied by \citet{Beauchamp1999}, who found that atmospheric parameters for DBVs determined assuming pure-He atmospheres were in some cases 3,000\,K hotter than those with $[\mathrm{H/He}]\,{=}\,\log(N_{\mathrm{H}}/N_{\mathrm{He}})\,{=}\,-3$. Subsequent studies that placed better observational constraints on [H/He] using spectroscopic H$\alpha$ coverage \citep{Voss2007,Bergeron2011,Rolland2018} found this systematic effect to be much smaller on average, of order a few hundred Kelvin. Regardless, in cases where only upper limits on [H/He] can be determined, a systematic uncertainty in \teff is present. It is also uncertain whether trace H should affect the location of the instability strip for DBAs. Some authors suggest the DBA instability should be a few hundred degrees cooler \citep{Fontaine2008}, and \citet{Bergeron2011} find the three DBAVs in their sample to pulsate at lower \teff relative to the DBVs, but \citet{vanGrootel_2017_1} found that trace H has no effect on the theoretical blue or red edges.

To address many of the issues mentioned above and use DBVs to their fullest potential, finding more DBVs (and NOVs) is the first step. The SDSS is a photometric and spectroscopic survey \citep{York2000,Eisenstein2011,Blanton2017} covering more than 10,000 deg$^2$ of the northern sky. To date, it is still the gold standard in terms of spectroscopic surveys, having increased the number of spectroscopically confirmed white dwarfs since the \citet{McCook1999} catalogue by more than an order of magnitude, from ${\sim}\,$3,000 to more than 30,000 \citep{Kleinman2004,Eisenstein2006,Kleinman2013,Kepler2015,Kepler2016,Kepler_2019_1}. Newer catalogues based on {\em Gaia} photometry have increased the number of white dwarfs by yet another order of magnitude \citep{GF2019,GF2021}, but without spectroscopic identifications, it is difficult to efficiently identify good DB candidates near the instability strip. Therefore, the SDSS is still one of the best resources available for increasing the number of known DBVs, allowing for more reliable identification of candidates for follow up time series photometry.

In this work, we report on an effort to identify DBVs using time series photometry from McDonald Observatory. Similar to \citet{Nitta_2009_1}, we identify candidate pulsating DBs and DBAs using atmospheric parameters determined from the SDSS DR10, DR12, and DR14 catalogues \citep{KK2015,Kepler_2019_1}. In Section~\ref{sec:observations} we discuss our target selection process and accompanying McDonald time series observations, in Section~\ref{sec:newDBVs} we report on the new DBVs we discovered, their detected periods, and the variability limits we place on NOVs, in Section~\ref{sec:DBstrip} we discuss the updated DB instability strip, in Section~\ref{sec:properties} we discuss the ensemble pulsation properties of the DBVs, and in Section~\ref{sec:conclude} we provide concluding remarks.

\section{Observations} \label{sec:observations}

\subsection{Target Selection} \label{sec:targets}

\defcitealias{KK2015}{KK15}
\defcitealias{Genest2019_1}{GBB19}

\begin{figure}[t!]
	\epsscale{1.18}
	\plotone{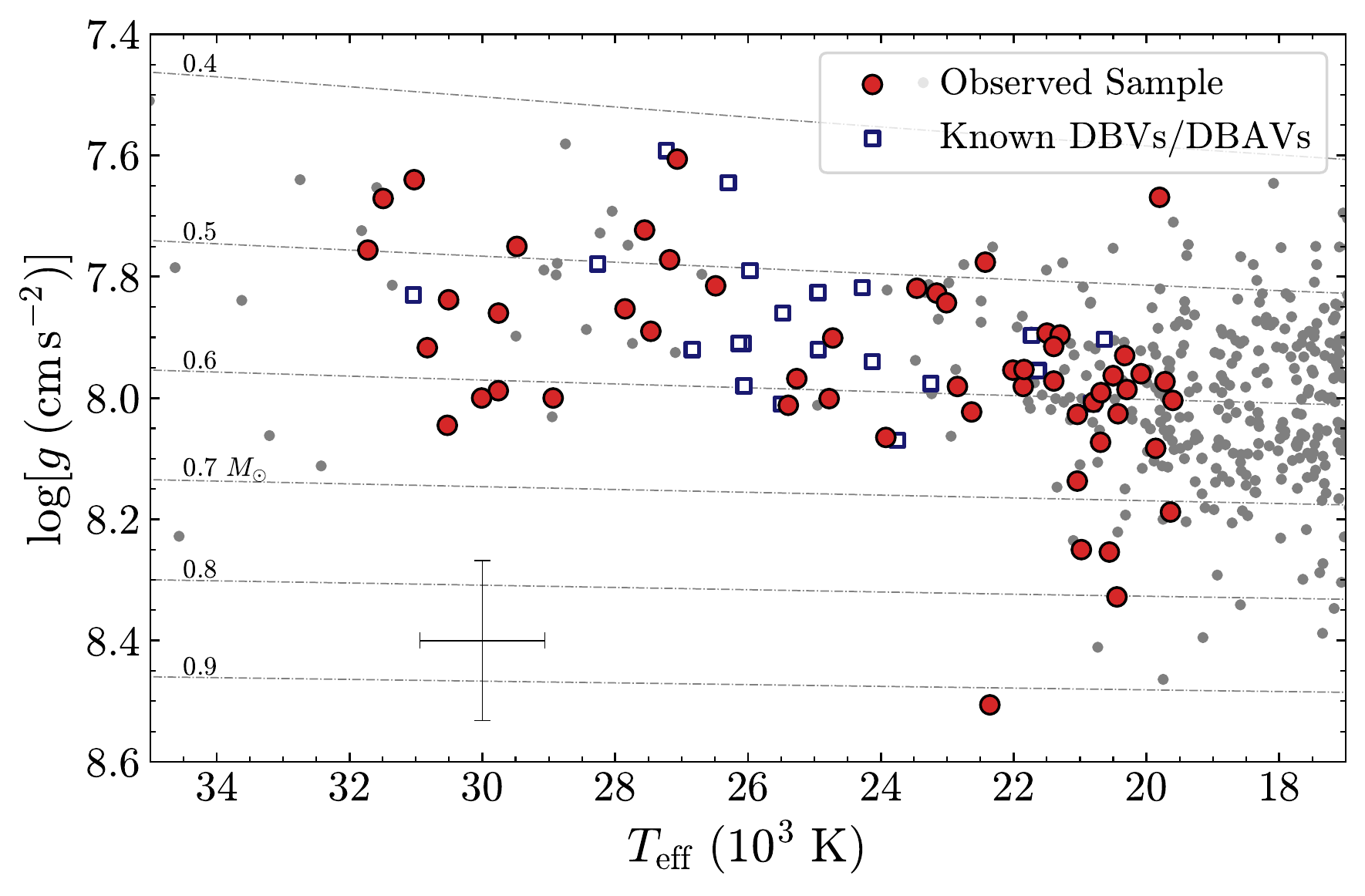}
	\caption{The sample of 55 DB/DBAs observed in this work (red circles) compared to the full sample of DB/DBA white dwarfs from \citet{KK2015} (grey circles) in the $\log(g)$ versus $T_{\mathrm{eff}}$ plane. Previously identified DBVs are marked with blue open squares, while grey dash-dotted lines show evolutionary tracks for white dwarfs of various masses with thin hydrogen envelopes ($M_{\mathrm{H}}/M_{\star}=10^{-10}$, \citealt{Bedard2020}). A typical error bar size for objects between 22,000 and 30,000\,K is shown in the bottom left. \label{fig:sample}}
\end{figure}

We selected our targets using the catalogue of spectroscopically confirmed DB white dwarfs identified in SDSS data releases 10 and 12 \citep[][hereafter KK15]{KK2015}, and data release 14 \citep{Kepler_2019_1}. In an effort to better define the edges of the instability strip, and because DB spectroscopic temperatures are often highly uncertain, we extended our search a few thousand degrees hotter and cooler with respect to both the empirical temperature range for DBVs and theoretical blue and red edges of \citet{vanGrootel_2017_1}. Ultimately, our search covered DB white dwarfs with spectroscopic effective temperatures between roughly 19,000 and 35,000\,K, including some DBAs. In Figure~\ref{fig:sample}, we show the effective temperatures and surface gravities of our sample of observed objects, 55 total, compared to the sample of previously identified DBVs and the DBs from \citetalias{KK2015}.

Based on previous efforts to discover new DBVs in the SDSS \citep[e.g.,][]{Nitta_2009_1}, we expected a large fraction of our targets to result in null detections of pulsations, i.e., not-observed-to-vary (NOV). For instance, just ${\approx}\,30\%$ of the DB white dwarfs observed by \citet{Nitta_2009_1} were found to pulsate. This low rate of detection is partly influenced by unreliable temperature determinations, such that objects whose temperatures place them within the instability may in fact lie outside the instability strip.

Still, the null detections of pulsations for objects near the blue or red edge place useful constraints on the boundaries of the DB instability strip, so we aimed to place limits on variability of $<5$\,mma (0.5\,\%) for all objects. We also aimed to observe all objects on two separate nights to minimize the chance that we observed a pulsating DB during a phase where destructive beating between two or more pulsation modes reduced the photometric variability below our detection threshold. We were able to obtain two nights of photometry for 48 out of 55 objects in our sample. Lastly, we mostly observed DB white dwarfs with $g$-band magnitudes $\lesssim19$ so we could place stronger NOV limits on null detections. The average SDSS $g$-band magnitude of our sample is about 17.9, with the full sample covering 15.7 to 19.3 mag. Only three objects out of 55 had magnitudes fainter than $g=19$.

\begin{figure*}[t!]
	\epsscale{1.18}
	\plotone{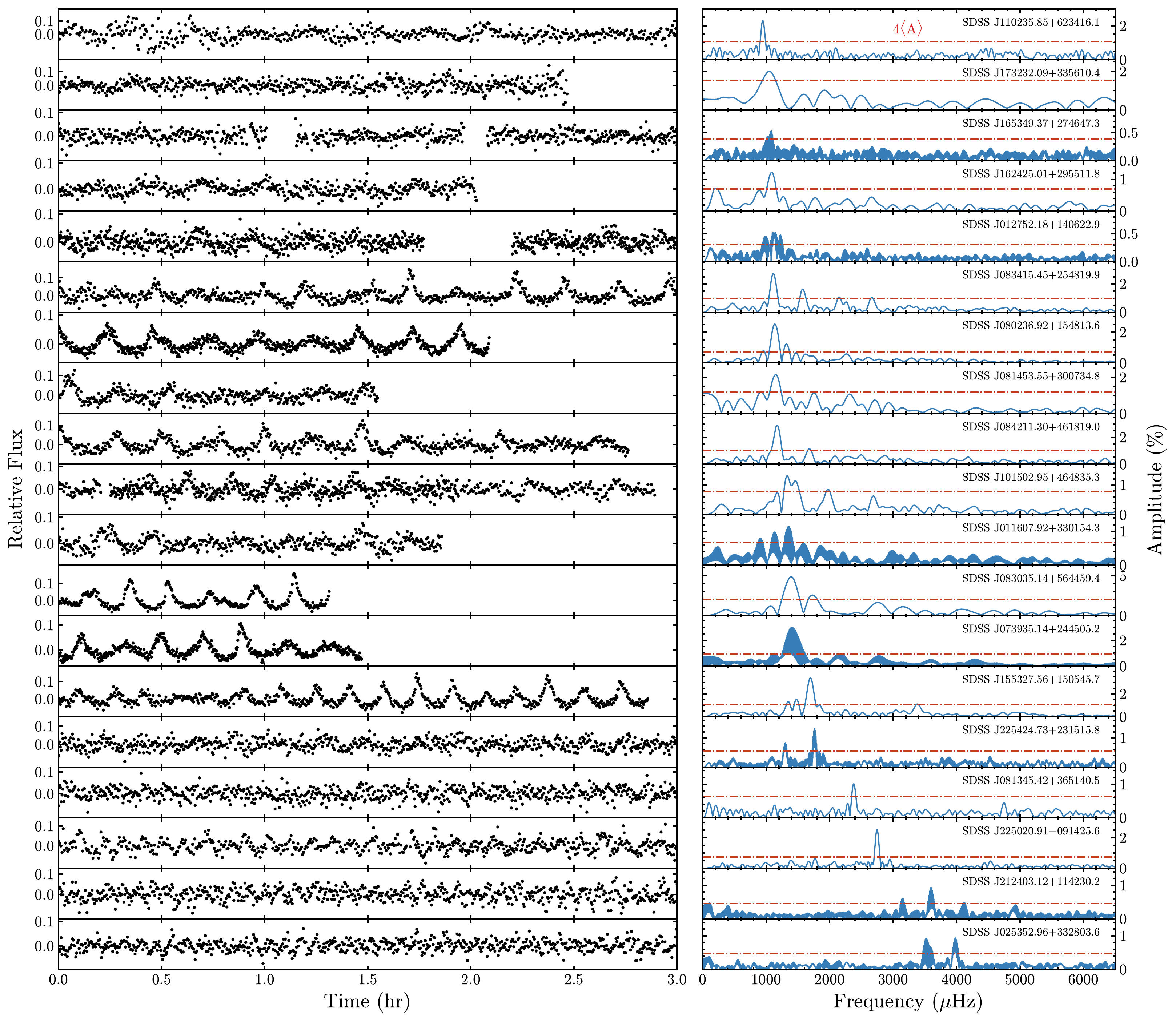}
	\caption{The light curves ({\em left}) and FTs ({\em right}) for each new DBV. Objects are ordered from top to bottom by increasing frequency (decreasing period) of the dominant mode observed. The \fourA significance threshold for each FT is shown with a horizontal red line. Some of the FTs are for multiple consecutive nights of observations, hence the higher density of peaks. Gaps in some light curves are caused by inclement weather. \label{fig:phot}}
\end{figure*}

\subsection{McDonald Photometry} \label{sec:photometry}

We carried out time series photometry on all of our targets using a Princeton Instruments ProEM frame-transfer CCD attached to the McDonald Observatory 2.1-m Otto Struve Telescope. We observed exclusively with a blue-bandpass BG40 filter which covers wavelengths between 3500 and 6500~\AA. Exposure times range from 3 to 30\,s depending on object brightness and weather conditions, with individual runs averaging 2.4\,hr in length. A full summary of our observations can be found in Appendix~\ref{appendix} in Table~\ref{tab:obs}.

Using standard calibration frames taken during each night of observations, we dark and flat-field corrected our images using the {\sc iraf} reduction suite. We then performed circular aperture photometry using the {\sc iraf} {\tt ccd\_hsp} package \citep{Kanaan2002} with aperture radii ranging from 2 to 10 pixels in half-pixel steps. Local sky subtraction was performed for each object using an annulus centered on each aperture.

We generated divided light curves using the {\sc Wqed} reduction software \citep{Thompson2013}, selecting the aperture size which maximized the S/N of the light curve. We removed any long term trends from the divided light curve with a low-order polynomial of degree two or less, and clipped any outliers or heavily cloud-affected data. Lastly, we used {\sc Wqed} to perform a barycentric correction to the mid-exposure time stamps of our images.


\begin{deluxetable*}{lrrrc|lrrrc}
\tablecaption{Observed Periods and Amplitudes for New DBVs  \label{tab:periods}}
\tabletypesize{\scriptsize}
\tablecolumns{10}
\tablewidth{0pt}
\tablehead{
\colhead{Name} & \colhead{Frequency} & \colhead{Period} & \colhead{Amp.} & \colhead{Mode ID} & 
\colhead{Name} & \colhead{Frequency} & \colhead{Period} & \colhead{Amp.} & \colhead{Mode ID} \\ [-0.2cm]
\colhead{(SDSS J)} & \colhead{($\mu$Hz)} & \colhead{(s)} & \colhead{(mma)} & \colhead{} &
\colhead{(SDSS J)} & \colhead{($\mu$Hz)} & \colhead{(s)} & \colhead{(mma)} & \colhead{}
}
\startdata
 011607.92$+$330154.3$^{\dagger}$ & 1357.2[0.6]  &  736.8[0.3]  & 10.3[1.1] & $f_{1a}$          & 083415.45$+$254819.9 & 1113.0[2.2]  &  898.5[1.8]  & 27.5[1.2] & $f_{9a}$  \\
                                  &  892.4[0.9]  & 1120.5[1.1]  &  7.5[1.1] & $f_{1b}$          &                      & 1576.7[3.7]  &  634.2[1.5]  & 16.3[1.2] & $f_{9b}$  \\
                                  & 1147.9[0.9]  &  871.1[0.7]  &  6.9[1.1] & $f_{1c}$          &                      & 2672.7[5.7]  &  374.2[0.8]  & 10.5[1.2] & $f_{9a}{+}f_{9b}$ \\
 012752.18$+$140622.9$^{\dagger}$ & 1107.2[0.7]  &  903.1[0.5]  &  4.7[0.6] & $f_{2a}$          &                      & 2144.0[6.1]  &  466.4[1.3]  &  9.8[1.2] & $2f_{9a}$    \\
                                  & 1137.9[0.6]  &  878.8[0.4]  &  5.6[0.6] & $f_{2b}$          & 084211.30$+$461819.0 & 1178.4[2.4]  &  848.6[1.7]  & 28.1[1.2] & $f_{10a}$  \\
                                  &  979.2[0.8]  & 1021.2[0.8]  &  4.0[0.6] & $f_{2c}$          &                      & 1669.5[5.8]  &  599.0[2.1]  & 11.6[1.2] & $f_{10b}$  \\
                                  & 1229.0[0.8]  &  813.7[0.5]  &  3.8[0.6] & $f_{2d}$          &                      & 1100.8[5.9]  &  908.5[4.8]  & 11.5[1.2] & $f_{10c}$  \\
 025352.96$+$332803.6$^{\dagger}$ & 3978.0[0.4]  & 251.38[0.03] &  9.4[0.7] & $f_{3a}$          & 101502.95$+$464835.3 & 1325.5[4.5]  &  754.4[2.6]  & 12.1[1.0] & $f_{11a}$  \\
                                  & 3519.4[0.5]  & 284.14[0.04] &  8.3[0.7] & $f_{3b}$          &                      & 1474.8[5.0]  &  678.1[2.3]  & 11.0[1.0] & $f_{11b}$  \\
                                  & 3588.8[0.7]  & 278.64[0.06] &  5.5[0.7] & $f_{3c}$          &                      & 1983.1[6.5]  &  504.3[1.7]  &  8.4[1.0] & $f_{11c}$  \\
 073935.14$+$244505.2$^{\dagger}$ & 1410.1[0.2]  &  709.2[0.1]  & 31.2[0.8] & $f_{4a}$          &                      & 1166.8[7.9]  &  857.1[5.8]  &  6.9[1.0] & $f_{11d}$  \\
                                  & 1174.8[0.5]  &  851.2[0.4]  &  9.6[0.8] & $f_{4b}$          &                      & 2688.0[8.5]  &  372.0[1.2]  &  6.5[1.0] & $2f_{11a}$    \\
                                  & 2152.7[0.5]  &  464.5[0.1]  & 10.7[0.8] & $f_{4c}$          & 110235.85$+$623416.1 &  946.1[3.3]  & 1057.0[3.7]  & 22.9[2.0] & $f_{12a}$  \\
                                  & 1630.6[0.6]  &  613.3[0.2]  &  8.5[0.8] & $f_{4d}$          & 155327.56$+$150545.7 & 1663.1[2.5]  &  601.3[0.9]  & 22.9[1.1] & $f_{13a}$  \\
                                  & 2735.5[0.6]  &  365.6[0.1]  &  7.9[0.8] & $2f_{4a}$         &                      & 1500.7[4.4]  &  666.4[2.0]  & 12.7[1.1] & $f_{13b}$  \\
                                  & 3597.4[0.9]  &  278.0[0.1]  &  5.5[0.8] & $f_{4a}{+}f_{4c}$ &                      & 3382.6[4.9]  &  295.6[0.4]  & 11.4[1.1] & $2f_{13a}$    \\
                                  & 1773.9[0.8]  &  563.7[0.2]  &  6.6[0.8] & $f_{4e}$          &                      & 1722.9[2.5]  &  580.4[0.8]  & 22.4[1.1] & $f_{13c}$  \\
                                  & 3004.1[1.0]  &  332.9[0.1]  &  5.1[0.8] & $f_{4a}{+}f_{4d}$ &                      & 1329.2[6.2]  &  752.4[3.5]  &  9.1[1.1] & $f_{13d}$  \\
                                  & 4273.1[1.1]  &  234.0[0.1]  &  4.7[0.8] & $3f_{4a}$         &                      & 1862.8[5.9]  &  536.8[1.7]  &  9.5[1.1] & $f_{13e}$  \\
 080236.92$+$154813.6             & 1166.4[1.3]  &  857.4[0.9]  & 35.5[0.6] & $f_{5a}$          & 162425.01$+$295511.8 & 1086.5[7.4]  &  920.4[6.3]  & 12.1[1.2] & $f_{14a}$  \\
                                  & 1227.2[2.4]  &  814.9[1.6]  & 18.9[0.6] & $f_{5b}$          & 165349.37$+$274647.3$^{\dagger}$ & 1078.3[0.4]  &  927.4[0.3]  &  5.4[0.7] & $f_{15a}$  \\
                                  & 2257.1[7.2]  &  443.0[1.4]  &  6.3[0.6] & $2f_{5a}$         & 173232.09$+$335610.4 & 1049.3[18.7] &  953.0[17.0] & 20.2[2.8] & $f_{16a}$  \\
                                  & 1694.1[9.5]  &  590.3[3.3]  &  4.8[0.6] & $f_{5c}$          & 212403.12$+$114230.2$^{\dagger}$ & 3596.3[0.2]  & 278.06[0.01] & 12.7[0.8] & $f_{17a}$  \\
 081345.42$+$365140.5             & 2378.9[4.0]  &  420.4[0.7]  & 10.1[1.1] & $f_{6a}$          &                      & 3602.1[0.3]  & 277.62[0.02] &  8.6[0.8] & $f_{17b}$  \\
                                  & 4748.9[8.9]  &  210.6[0.4]  &  4.5[1.1] & $2f_{6a}$         &                      & 3157.8[0.4]  & 316.67[0.04] &  5.9[0.8] & $f_{17c}$  \\
 081453.55$+$300734.8             & 1150.5[8.2]  &  869.2[6.2]  & 21.4[1.8] & $f_{7a}$          &                      & 4111.1[0.5]  & 243.24[0.03] &  5.2[0.8] & $f_{17d}$  \\
 083035.14$+$564459.4             & 1384.7[2.5]  &  722.2[1.3]  & 47.7[1.0] & $f_{8a}$          &                      & 4933.3[0.6]  & 202.71[0.02] &  4.2[0.8] & $f_{17e}$  \\
                                  & 1763.9[5.4]  &  566.9[1.7]  & 21.8[1.0] & $f_{8b}$          & 225020.91$-$091425.6 & 2747.1[1.4]  &  364.0[0.2]  & 25.3[1.2] & $f_{18a}$  \\
                                  & 2761.7[6.4]  &  362.1[0.8]  & 18.3[1.0] & $2f_{8a}$         & 225424.73$+$231515.8$^{\dagger}$ & 1761.3[0.2]  &  567.8[0.1]  & 13.6[0.9] & $f_{19a}$  \\
                                  & 3160.0[9.8]  &  316.5[1.0]  & 12.0[1.0] & $f_{8a}{+}f_{8b}$ &                      & 1301.0[0.3]  &  768.6[0.2]  &  8.9[0.9] & $f_{19b}$  \\
                                  & 4127.3[12.6] &  242.3[0.7]  &  9.3[1.0] & $3f_{8a}$         &                      &              &              &           &      \\
                                  & 4553.3[16.5] &  219.6[0.8]  &  7.1[1.0] & $2f_{8a}{+}f_{8b}$&                      &              &              &           &      \\
                                  & 5486.5[18.5] &  182.3[0.6]  &  6.3[1.0] & $4f_{8a}$         &                      &              &              &           &      \\
\enddata
\tablecomments{Uncertainties are given in brackets next to each value. For combination modes, multiple possibilities often exist within the frequency resolution of our light curves. We list here the option we consider most likely based on the frequency match and parent amplitudes.}
\tablenotetext{\dagger}{Indicates objects that were observed on two or more consecutive nights.}
\end{deluxetable*}

\section{New DBVs and Limits on Variability} \label{sec:newDBVs}

We observed a total of 55 DBs and DBAs with the McDonald 2.1-m telescope over the course of 270 total hours. We detected pulsations in 19 new DBVs, a success rate of about 35\%, similar to that of \citet{Nitta_2009_1}. In Figure \ref{fig:phot}, we present the light curve for each newly discovered DBV, along with the associated Fourier transform (FT) calculated using the {\sc Period04} program \citep{Period04}, ordered from top to bottom by increasing frequency of the dominant mode. In this section we characterize the DBVs, measuring their pulsation periods and amplitudes, and also place limits on variability for those objects which show no evidence for pulsations. 

To identify significant pulsation modes, we first calculate FTs of our light curves using {\sc Period04}, oversampling the frequencies by a factor of 20. We then adopt an iterative ``prewhitening'' procedure similar to that described in \citet{Bell2017a}, first assessing whether the highest peak in the FT rises above a \fourA threshold, where $\langle$A$\rangle$ is the average amplitude of the FT between 500 and 10,000~$\mu$Hz (\fourA approximates a 0.1\% false alarm probability level for relatively short time series observations; \citealt{Breger_1993_1,Kuschnig_1997_1}). If the peak is significant, we perform a non-linear least squares fit of a sinusoid to the light curve, using the peak frequency and amplitude as initial guesses. We then prewhiten the light curve by subtracting the best-fit sinusoid and calculating an FT of the residuals. We find the highest peak remaining and repeat the above process until no significant peaks remain, each time fitting a sum of sinusoids to the light curve.

For objects with multiple nights of observations, the gaps between runs introduce cycle count ambiguities which manifest as complex aliasing structures in the spectral window of the FT. These aliasing structures make it difficult to identify the correct periods and allow for multiple viable period solutions \citep[e.g.,][]{Bell2017b,Bell2018}. To reduce the impact of aliasing on our determined periods, we only combine light curves for seven objects that were observed on consecutive nights, and otherwise use the highest quality single-night light curve to characterize the periods. We only determine one period solution per object and report analytical least-squares uncertainties \citep{Montgomery1999} for our frequencies, periods, and amplitudes. We stress, however, that for the seven objects with combined multi-night light curves, the characteristic spacing between aliases of $\approx$11.6~$\mu$Hz (the inverse of one Sidereal day) sets extrinsic errors on our frequencies that are much larger than the reported formal uncertainties. 

\begin{deluxetable}{lcc}
\tablecaption{References for the Previously Known DBVs \label{tab:known_DBVs}}
\tablecolumns{3}
\tabletypesize{\scriptsize}
\tablehead{
\colhead{Name}     & \colhead{Spectroscopy} & \colhead{Pulsation Periods}
}
\startdata
 SDSS\,J034153.03$-$054905.9   & \citealt{KK2015}           & \citealt{Nitta_2009_1} \\
 SDSS\,J094749.40$+$015501.9   & \citealt{Kleinman2013}     & \citealt{Nitta_2009_1} \\
 SDSS\,J104318.45$+$415412.5   & \citealt{Kleinman2013}     & \citealt{Nitta_2009_1} \\
 SDSS\,J122314.25$+$435009.1   & \citealt{KK2015}           & \citealt{Nitta_2009_1} \\
 SDSS\,J125759.04$-$021313.4   & \citealt{Kepler_2019_1}    & \citealt{Nitta_2009_1} \\
 SDSS\,J130516.51$+$405640.8   & \citealt{KK2015}           & \citealt{Nitta_2009_1} \\
 SDSS\,J130742.43$+$622956.8   & \citealt{KK2015}           & \citealt{Nitta_2009_1} \\
 SDSS\,J140814.64$+$003839.0   & \citealt{Kepler_2019_1}    & \citealt{Nitta_2009_1} \\
 EC\,01585$-$1600              & \citealt{Rolland2018}      & \citealt{Bell2019}     \\
 EC\,04207$-$4748              & \citealt{Koester2014_2}    & \citealt{Kilkenny2009} \\
 EC\,05221$-$4725              & ---                        & \citealt{Kilkenny2009} \\
 KUV\,05134$+$2605             & \citealt{Rolland2018}      & \citealt{Bognar2014}   \\
 CBS\,114                      & \citealt{Rolland2018}      & \citealt{Handler2002}  \\
 PG\,1115$+$158                & \citealt{Rolland2018}      & \citealt{Winget1987}   \\
 PG\,1351$+$489                & \citealt{Rolland2018}      & \citealt{Radaelli2011} \\
 PG\,1456$+$103                & \citealt{Rolland2018}      & \citealt{Handler2002}  \\
 GD\,358                       & \citealt{Rolland2018}      & \citealt{Provencal2009} \\
 PG\,1654$+$160                & \citealt{Rolland2018}      & \citealt{Handler2003}  \\
 PG\,2246$+$121                & \citealt{Rolland2018}      & \citealt{Handler2001}  \\
 EC\,20058-5234                & \citealt{Koester2014_2}    & \citealt{Sullivan2008} \\
 PG\,0112$+$104                & \citealt{Rolland2018}      & \citealt{Hermes_2017_2} \\
 KIC\,8626021                  & \citealt{Giammichele2018}  & \citealt{Ostensen2011} \\
 {\bf EPIC\,228782059}         & \citealt{Kepler_2019_1}    & \citealt{Duan2021}     \\
 SDSS\,J085202.44$+$213036.5   & \citealt{KK2015}           & \citealt{Nitta_2009_1} \\
 WD\,J025121.71$-$125244.85    & ---                        & \citealt{Rowan2019}    \\
 SDSS\,J102106.69$+$082724.8   & \citealt{KK2015}           & \citealt{Rowan2019}    \\
 SDSS\,J123654.96$+$170918.7   & \citealt{KK2015}           & \citealt{Rowan2019}    \\
 {\bf WD\,J132952.63$+$392150.8}     & ---                        & \citealt{Rowan2019}    \\
\enddata
\end{deluxetable}

In Table~\ref{tab:periods}, we list the identified significant periods for each new DBV and assign them a preliminary mode ID, $f_{ij}$, where $i$ is a unique number for each object, and $j$ is unique letter for each mode within the object (e.g. $f_{1a}$). We generate all possible additive combinations of the significant periods for each object, including harmonics. If any combinations match with a significant frequency within the frequency resolution defined by the length of the light curve, we identify them as possible combination modes. Due to the low frequency resolutions of our light curves, multiple combination possibilities often exist for many modes, and some might actually be independent modes, so we do not constrain any periods to exact arithmetic relationships when fitting with {\sc Period04}. In the mode ID column of Table~\ref{tab:periods}, we list only the combination we consider most likely, which is typically the combination involving the closest frequency match or the highest amplitude parent modes. 

For objects with multiple nights of data, the frequency resolution of the combined light curve is often much smaller than the width of the spectral window, so in these cases we use the frequency resolution of the longest single-night light curve as the matching tolerance when searching for combinations. We also note that even for modes which are unlikely to be combinations, our observing runs are too short in most cases to resolve closely spaced modes or rotationally split multiplets, and more extensive observing would be needed to properly identify the independent pulsation modes in these objects. Still, throughout the analysis in this work, we treat any significant frequencies that are unlikely to be combinations as independent modes in these stars.

For objects that show no significant peaks during any of the nights observed, we classify them as not-observed-to-vary (NOV). We identified 36 new NOVs, and use the \fourA thresholds of their FTs to place limits on their variability. 34 out of 36 objects (94\%) were observed on at least two separate occasions. For these objects, we calculate NOV limits using their combined light curves, regardless of how many nights separated the observations. For 32 objects, 89\% of our sample, we achieve NOV limits less than 0.5\%, and reach an average limit of 3.3\,mma for our full sample. We list the NOV limits along with the number of observing runs acquired for each object in the rightmost columns of Table~\ref{tab:NOV_Params} located in Appendix~\ref{appendix}. 

White dwarfs are known to exhibit pulsation amplitudes below our NOV limits, and it is also possible we observed some objects only during phases of destructive beating between pulsation modes. Additional observations may be needed for some objects to rule out pulsations with higher confidence, especially for DBs with \teff near the blue edge where pulsation amplitudes tend to be small. Still, our NOV limits represent an improved assessment of variability throughout the DB instability strip compared to \citet{Nitta_2009_1}, where only 15\% of NOVs were observed on more than one night, and only 30\% have NOV limits ${<}5\,$mma, giving an average NOV limit of 8.9\,mma. With our observations, we were also able to identify pulsations in one DB, SDSS\,J101502.95+464835.3, which was previously identified as an NOV by \citet{Nitta_2009_1} with a variability limit of 7.2\,mma using a single 1.7\,hr run. We detected four independent pulsation modes in this object with amplitudes of 6.9$-$12.1\,mma, indicating destructive beating was likely occurring during the observations of \citet{Nitta_2009_1}, and highlighting the need for more extensive observations to improve assessment of variability throughout the DB instability strip.

\section{The DB/DBA Instability Strip} \label{sec:DBstrip}

In this section we use our sample of new DBVs and NOVs alongside previously known DBVs to provide new constraints on the empirical limits of the DB/DBA instability strip, and compare with the most recent theoretical calculations. For the new DBVs and NOVs in our sample, we took their spectroscopic \teff and $\log(g)$ values from \citetalias{KK2015} and \citet{Kepler_2019_1}, who fit 1D white dwarf atmospheric models to SDSS spectra from DR10, DR12, and DR14. Their models employ the ML2/$\alpha$ version \citep{Bohm71,Tassoul1990} of the Mixing Length Theory \citep[MLT,][]{Bohm1958} to describe convective energy transport, with the mixing length parameter $\alpha$ set to 1.25 as calibrated by \citet{Beauchamp1999} and \citet{Bergeron2011}. 

\begin{figure*}[t!]
	\epsscale{0.8}
	\plotone{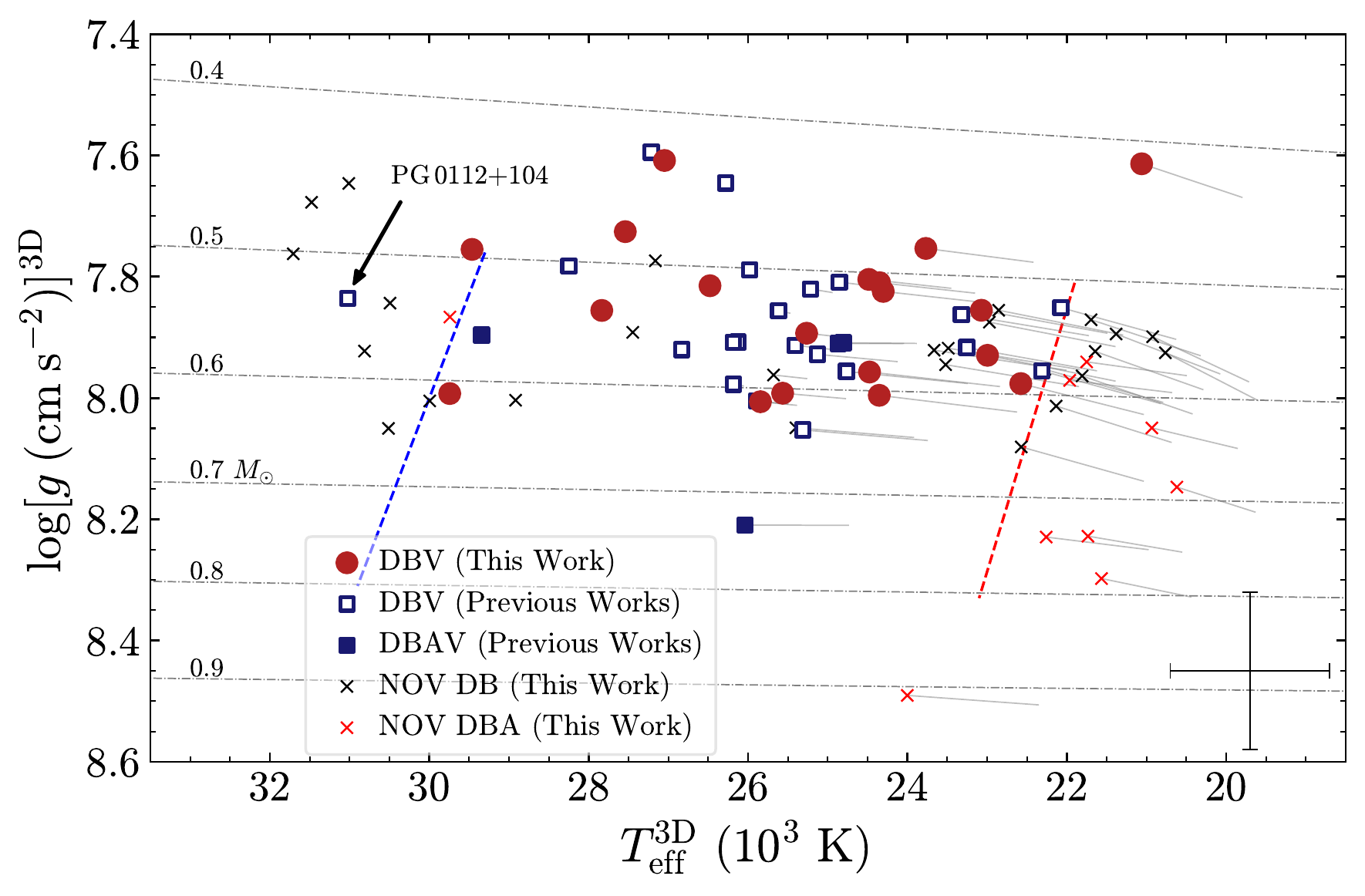}
	\caption{The DB/DBA instability strip with 3D $T_{\mathrm{eff}}$ and $\log(g)$ corrections from \citet{Cukanovaite2021} applied. Lines drawn from each point indicate the change from 1D spectroscopic $T_{\mathrm{eff}}$ and $\log(g)$ values, and a typical error bar size is shown in the bottom right corner. In comparison with the theoretical blue and red edges from \citet{vanGrootel_2017_1} (blue and red dashed lines, respectively), the observed blue edge defined by PG\,0112+104 is still significantly hotter, while the observed red edge appears just slightly hotter. As shown in Figure~\ref{fig:tlogg_comp}, however, the observed red edge can move significantly between studies using different model atmospheres and fitting methods. \label{fig:strip}}
\end{figure*}

\begin{figure*}[t!]
	\epsscale{0.85}
	\plotone{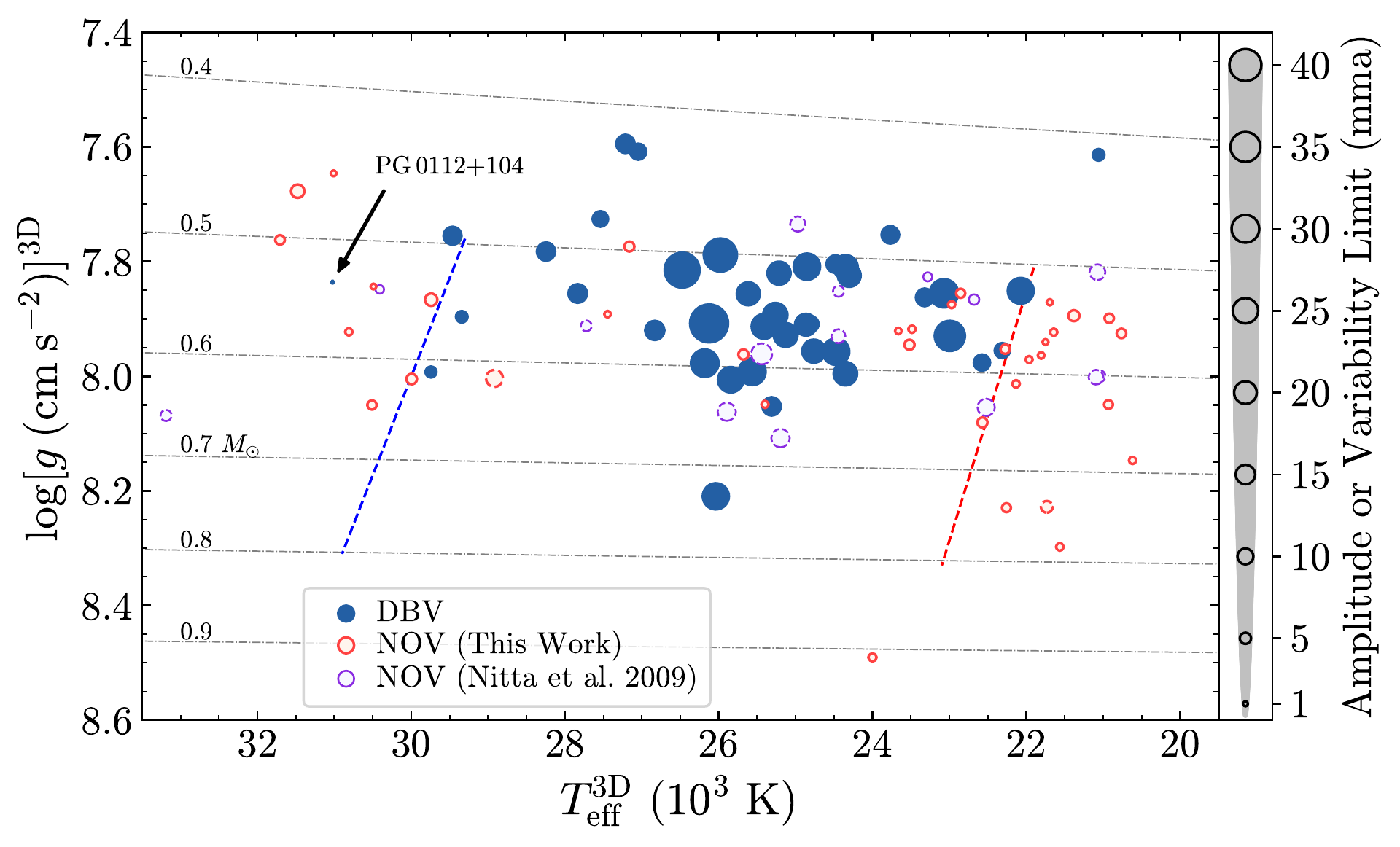}
	\caption{The same DB/DBA instability strip as Figure~\ref{fig:strip}, but with DBVs (blue circles) sized according to the amplitude of their dominant mode, and NOVs (open red and purple circles) sized according to their variability limit. Open red circles are new NOVs from this work, while open purple circles are NOVs from \citet{Nitta_2009_1}. Dashed versus solid marker edges for NOVs indicate whether they were observed on one or more than one night, respectively. Our NOV limits are generally much lower than the typical pulsation amplitudes of DBVs, except near the blue edge where the measured pulsation amplitudes are often $\lesssim5\,$mma (e.g., the highest amplitude mode in PG\,0112+104 is ${<}0.3\,$mma, \citealt{Hermes_2017_2}). Four of the new NOVs in our sample have \teff and \logg more than 1$\sigma$ inside the theoretical instability strip, all of which have two or more nights of observations and variability limits ${<}5\,$mma (see Table~\ref{tab:NOV_Params}). \label{fig:ampstrip}}
\end{figure*}

For the 28 previously known DBVs, only 10 are found in \citetalias{KK2015} or \citet{Kepler_2019_1}. For the remaining 18 objects, we acquire their spectroscopic \teff, \logg, and [H/He] parameters from various works in the literature that are listed in Table~\ref{tab:known_DBVs}, all of which use ML2/$\alpha=1.25$. Three of these 18 objects, EC\,05221$-$4725, WD\,J025121.71$-$125244.85, and WD\,J132952.63+392150.8 have not been spectroscopically analyzed yet, so we exclude them from our analysis. Thus, in the proceeding sections we analyze a total of 80 pulsating and NOV DB/DBAs.

The atmospheric models and fitting procedures differ in some ways between these works, but we opted for a complete rather than homogeneous sample so we could present a complete census of all currently known DBVs. Because of this inhomogeneity, some caution should be taken when interpreting the limits of the observed instability strip which can vary based on which atmospheric models and fitting methods are used. One consistent factor, however, is that all the atmospheric parameters used in this work were calculated using the ML2/$\alpha=1.25$ mixing-length prescription. This allows us to apply the 3D \teff and \logg corrections from \citet{Cukanovaite2021} later on to see how they might affect the empirical extent of the instability strip, and compare with spectroscopic and photometric parameters derived in separate works \citep[e.g.,][]{Genest2019_1,GF2021}.

With the exception of the two objects fit by \citet{Kleinman2013}, all the works mentioned above consider the possibility of trace hydrogen in the atmospheres of the DBs they studied, and use the detection or non-detection of H$\alpha$ to measure or place upper limits on [H/He] at the photosphere for each object. For all 19 new DBVs presented in this work, only upper limits on [H/He] have been determined, while nine of the 36 new NOVs have detected H making them DBAs. Only four of the previously known DBVs are classified as DBAs \citep{Giammichele2018,Rolland2018}.

We list the atmospheric properties for all new DBVs and NOVs in Appendix~\ref{appendix} in Tables~\ref{tab:DBV_Params} and \ref{tab:NOV_Params}, respectively. The \teff and \logg uncertainties for new DBVs and NOVs are the formal values from \citetalias{KK2015} and \citet{Kepler_2019_1}, with 3.1\% relative \teff and 0.12~dex \logg added in quadrature to account for external uncertainties estimated by \citetalias{KK2015}. These are similar to the uncertainties estimated by \citet{Genest2019_1} for their SDSS spectroscopic sample. For previously known DBVs we use uncertainties quoted by the same works where we obtained the atmospheric parameters.

Lastly, prior to investigating the observational extent of the DB/DBA instability strip, we apply the 3D \teff and \logg corrections calculated by \citet{Cukanovaite2021} to all of the new and previously known DBVs and NOVs. For DBs with only upper limits on [H/He], we perform the corrections using the lowest possible H-abundance in the \citet{Cukanovaite2021} grid, [H/He]=$-$10. Otherwise, we use the measured [H/He] values from the literature as input to the correction functions. The corrections for our sample of objects are almost negligible at temperatures above 26,000\,K, but below this temperature the corrections change the 1D \teff and \logg by about $+$1,200\,K and $-$0.03~dex on average, respectively.

In Figure~\ref{fig:strip}, we show the updated DB/DBA instability strip after the application of 3D corrections. To indicate the direction and magnitude of the 3D corrections, we draw vectors from each point back to their respective 1D \teff and \logg. The 3D corrections move nearly every DBV whose 1D atmospheric parameters are cooler than the theoretical red edge inside the theoretical instability strip of \citet{vanGrootel_2017_1}. Many NOVs move inside the instability strip as well, but most of them are still within 1$\sigma$ uncertainties of both the theoretical red edge and the coolest DBVs, and could conceivably be non-pulsators beyond the red edge. Overall, the 3D corrections appear to improve the agreement between the empirical and theoretical red edges, though potentially the theoretical red edge is still slightly too cool given that all the known cool DBVs are located inside the strip, and statistically there ought to be some NOVs beyond the theoretical red edge given their large temperature uncertainties. The eight coolest DBVs in our observed instability strip are on average 650\,K hotter than the theoretical red edge.

To date, PG\,0112+104 is still the hottest known DBV \citep{Hermes_2017_2} with \teff$=31{,}040\pm1060\,$K \citep{Rolland2018}, though we do find two new DBVs which are now the second and third hottest DBVs according to their spectroscopic \teff, SDSS\,J012752.18+140622.9 ($T_{\mathrm{eff}}^{\mathrm{3D}}\,{=}\,29{,}740\pm970\,$K) and SDSS\,J212403.12+114230.2 ($T_{\mathrm{eff}}^{\mathrm{3D}}\,{=}\,29{,}460\pm1140\,$K). Alongside KIC\,8626021 \citep{Bischoff-Kim2014,Giammichele2018}, these help place additional constraints on the blue (hot) edge of the instability strip. We do note, however, that the detected pulsation periods of SDSS\,J012752.18+140622.9 are relatively long, between 810 and 1020\,s, which are more typical of cool-edge pulsators and call into question the accuracy of the atmospheric parameters for this object. We discuss this object further in Section~\ref{sec:properties}. Meanwhile, PG\,0112+104 still poses a significant challenge to the current theoretical blue edge from \citet{vanGrootel_2017_1} who predict driving to begin around 29,500\,K at $\log(g)=7.8$. A higher convective efficiency (higher $\alpha$) is still required to bring the observed and theoretical blue edges into agreement.

\begin{figure*}[t!]
	\epsscale{0.9}
	\plotone{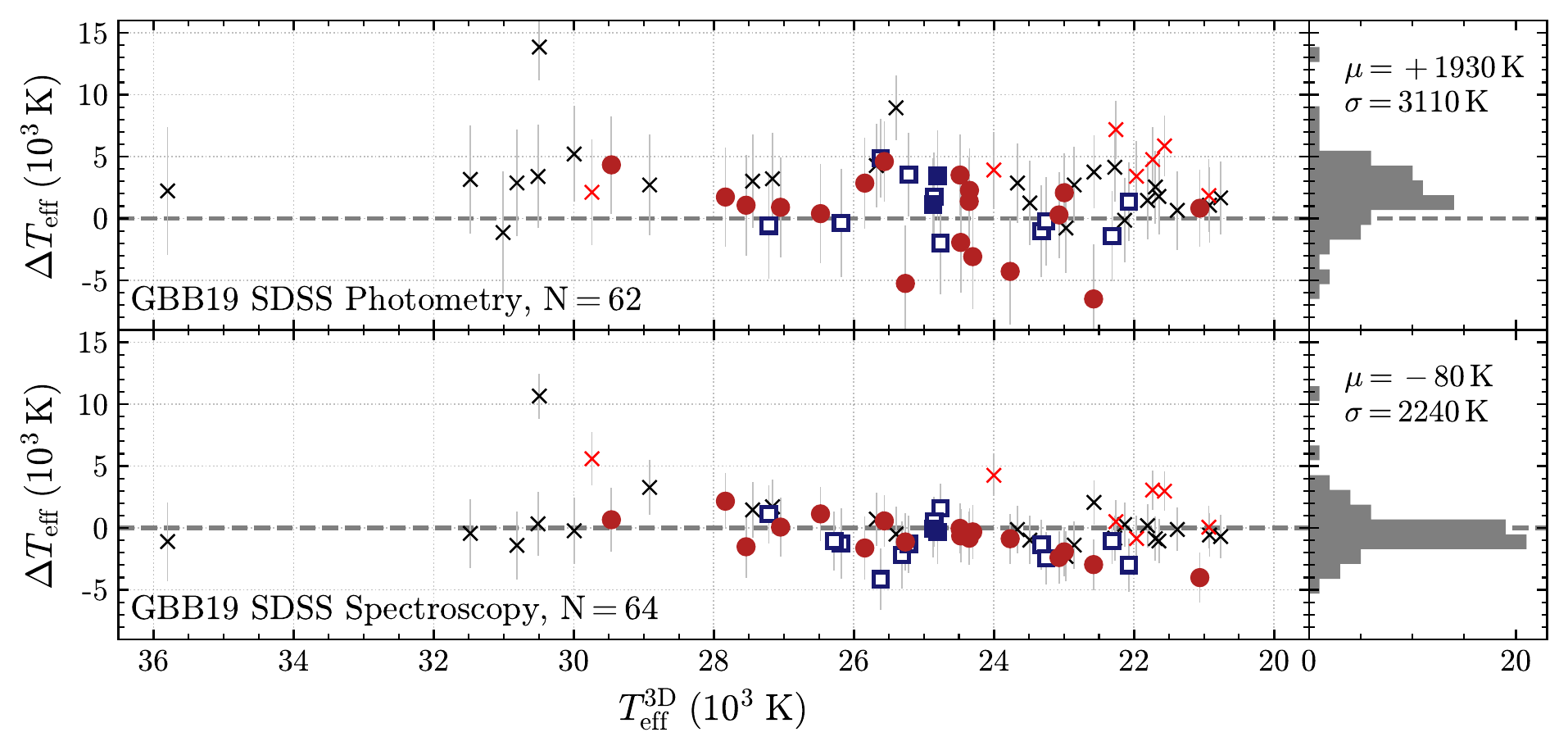}
	\caption{\teff differences between those used in this work (see Tables~\ref{tab:known_DBVs}, \ref{tab:DBV_Params}, and \ref{tab:NOV_Params}) and those provided by \citetalias{Genest2019_1} who use SDSS photometry ({\em top}, 62 objects in common) and spectroscopy ({\em bottom}, 64 objects in common). $\Delta T_{\mathrm{eff}} = T_{\mathrm{eff}}^{\mathrm{Our\,Sample}} - T_{\mathrm{eff}}^{\mathrm{GBB19}}$, and 3D corrections have been applied to all spectroscopic parameters in this figure. Histograms of the \teff differences are shown on the right along with their averages ($\mu$) and root-mean-square deviations ($\sigma$). While the average spectroscopic $\Delta$\teff for the full sample is negligible compared to uncertainties in \teff, the spectroscopic parameters from \citetalias{Genest2019_1} for DBVs below $24{,}000\,$K appear to be higher on average and would suggest a hotter red edge (see Figure~\ref{fig:strip_comparison}). The photometric parameters are systematically lower, but there is a larger amount of scatter. There are also a few NOVs within our instability strip which are moved outside the instability strip according to the \citetalias{Genest2019_1} spectroscopic or photometric parameters. \label{fig:tlogg_comp}}
\end{figure*}

\begin{figure*}[t!]
	\epsscale{1.15}
	\plotone{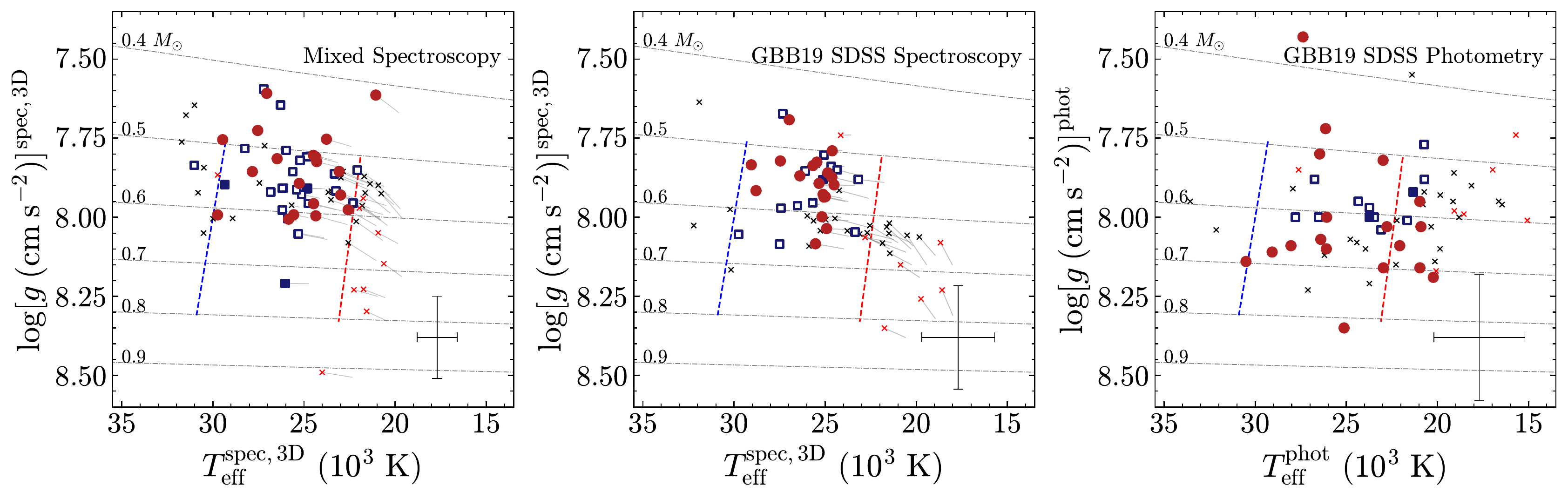}
	\caption{The instability strips generated using atmospheric parameters from different studies and fitting methods. On the left is the same instability strip from Figure~\ref{fig:strip}, which uses an inhomogeneous set of spectroscopic parameters from a wide range of studies (see Tables~\ref{tab:known_DBVs}, \ref{tab:DBV_Params}, and \ref{tab:NOV_Params}) but includes the largest number of known pulsating and NOV DBs and DBAs (80 objects). In the middle and right panels are instability strips generated from \citetalias{Genest2019_1} spectroscopic (64 objects) and photometric (62 objects) parameters, respectively, which rely solely on SDSS observations. The symbol shapes and colors are the same as Figure~\ref{fig:strip}, and we again use lines connected to each object with spectroscopic parameters to indicate the change from 1D \teff and \logg values. A typical error bar size is shown in the bottom right corner of each panel, with the photometric parameters having notably larger uncertainties. The observed red edge changes by about $\pm$2000\,K between different studies and fitting methods. The observed blue edge appears more stable but is harder to define given the small number of objects, especially since the hottest known DBV, PG\,0112+104, is only present in the left panel. \label{fig:strip_comparison}}
\end{figure*}

We also find several NOVs inside the instability strip. In Figure~\ref{fig:ampstrip}, we show the same instability strip as Figure~\ref{fig:strip} but size each DBV marker according to the amplitude of the dominant mode, and each NOV marker according to the variability limit. We include the \citet{Nitta_2009_1} sample of NOVs in this figure for a comparison of the variability limits of our samples, which is also described in text at the end of Section~\ref{sec:newDBVs}. As shown in Figure~\ref{fig:ampstrip}, the majority of our NOVs and some of the NOVs from \citet{Nitta_2009_1} have variability limits well below the typical measured pulsation amplitudes of DBVs at similar \teff and \logg, except near the blue edge where the measured pulsation amplitudes are often less than our typical detection thresholds of 2$-$5\,mma. Pulsation modes in PG\,0112+104, the hottest known DBV, reach only 0.3\,mma amplitude in the {\em Kepler} bandpass \citep{Hermes_2017_1}. 

Due to the large temperature uncertainties in these objects, some NOVs may in fact belong outside the instability strip, but some objects remain well within the instability strip even when using atmospheric parameters from different studies. In particular, there are four objects in our NOV sample, all with multiple nights of observations and NOV limits below 5\,mma, which reside near the middle of the instability strip and more than 1$\sigma$ away from both the blue and red theoretical edges. SDSS\,J081656.17+204946.0, for example, with an NOV limit from three nights of photometry of 1.73\,mma, has $T_{\mathrm{eff}}\,{=}\,27{,}400$ and $\log(g)\,{=}\,7.89$ \citepalias[][SDSS spectroscopy]{KK2015}, $T_{\mathrm{eff}}\,{=}\,25{,}700$ and $\log(g)\,{=}\,8.00$ \citep[][SDSS spectroscopy]{Genest2019_1}, and $T_{\mathrm{eff}}\,{=}\,24{,}400$ and $\log(g)\,{=}\,8.08$ \citep[][SDSS photometry]{Genest2019_1}, all of which place this object well within the instability strip. Most of the NOVs in our sample within the instability strip were observed on two separate nights, decreasing the chance we caught them during cycles of destructive beating between modes, but such an effect may still account for some NOVs inside the instability strip, especially those in the \citet{Nitta_2009_1} sample which in most cases have only one night of observations. To place stronger upper limits on variability, especially near the blue edge, more extensive time series photometry of these objects is required.

Even though we double the number of cool DBVs within 2,000\,K of the red edge from four to eight, the exact location of the observed red edge is still difficult to determine with our sample given the large temperature uncertainties and that the atmospheric parameters used here represent an inhomogeneous spectroscopic sample. Temperatures and surface gravities can vary by large amounts for single objects due to different atmospheric models, fitting routines, and properties of the observational data such as S/N, resolution, and wavelength coverage. To illustrate this, we show in Figure\,\ref{fig:tlogg_comp} the difference in \teff and \logg for our sample of DBVs and NOVs and those in common with the spectroscopic and photometric analyses of \citet{Genest2019_1}.

\citet{Genest2019_1} (abbreviated as \citetalias{Genest2019_1} hereafter), use both SDSS spectroscopy and photometry from DR12 and earlier to determine two independent sets of temperatures and surface gravities for their objects, one based on the spectroscopic method and the other based on the photometric method. Out of the 80 objects in our sample with prior spectroscopic observations, we find 64 objects in  \citetalias{Genest2019_1} with SDSS spectroscopy, and 62 objects with SDSS {\em ugriz} photometry. Unsurprisingly, given that most of our atmospheric parameters come from \citetalias{KK2015} who also use SDSS spectra from DR12 and below, the \teff and \logg we use in this work agree closely with the spectroscopic analysis of \citetalias{Genest2019_1}. 

For the 64 objects in \citetalias{Genest2019_1} with spectroscopic parameters, the \teff and \logg of our sample are $80\,$K cooler and $0.05\,$dex lower with respect to \citetalias{Genest2019_1} values, with root-mean-square (RMS) deviations of 2200\,K and 0.10\,dex. Despite the small average difference in \teff, the coolest DBVs in our sample with \teff$\lesssim$24,000\,K have systematically lower \teff with respect to the spectroscopic sample of \citetalias{Genest2019_1}, which would produce a red edge about 2000\,K hotter. We show this effect more clearly in the middle panel of Figure~\ref{fig:strip_comparison} which displays the DB instability strip as defined by the spectroscopic parameters from \citetalias{Genest2019_1}. After applying 3D-corrections only two DBVs have \teff less than 24,000\,K and the majority of DBVs are clumped around 25,000\,K. This clumping effect is primarily caused by the 3D spectroscopic corrections, but even without applying 3D corrections the coolest DBVs in \citetalias{Genest2019_1} would still be systematically hotter relative to our inhomogeneous sample.

For the 62 objects in \citetalias{Genest2019_1} with photometric parameters, the average \teff for our 3D-corrected spectroscopic sample is actually 1900\,K hotter with respect to \citetalias{Genest2019_1} values, though the amount of scatter is significantly larger with RMS deviations around 3100\,K. This effect can again be seen more clearly in the right panel of Figure~\ref{fig:strip_comparison}, where a much larger number of DBVs and DBAVs are seen at temperatures cooler than the proposed theoretical red edge of \citet{vanGrootel_2017_1} than in either of the spectroscopic samples. In this case, though, the difference between our sample of spectroscopic parameters and the photometric parameters of \citetalias{Genest2019_1} near the red edge would actually be greatly reduced without the application of 3D spectroscopic corrections, although the photometric parameters are still about 1000\,K cooler on average for the whole sample.

As seen in Figure~\ref{fig:strip_comparison}, while the red edge can vary by about $\pm2000\,$K between the different studies and fitting methods, the effect on the blue edge is much harder to determine due to the low number of hot DBVs. It appears to be somewhat more stable than the red edge, but unfortunately the hottest known DBV, PG\,0112+104, does not fall within the SDSS footprint and so is not contained in the \citetalias{Genest2019_1} sample. At least one hot DBV is always found near to the theoretical blue edge between the different studies and methods, though perhaps a more systematic drop in \teff can be seen in the photometric sample relative to the spectroscopic sample.

We also attempted to compare the atmospheric parameters for our sample of DBVs and NOVs with those determined by \citet{GF2021}. They use the photometric method with {\em Gaia} eDR3 photometry and parallax to derive \teff and \logg assuming either a pure-H, pure-He, or mixed H-He atmosphere with [H/He]=$-$5. The {\em Gaia} photometry, however, is not well suited for measuring the temperatures of hot objects given its broad and relatively red passbands, with average uncertainties of about 4500\,K and 0.26~dex in \teff and \logg, respectively, for our sample of objects. These uncertainties prevent any meaningful comparison between the atmospheric parameters of our sample, but at the very least we do not find any significant systematic differences in \teff or \logg in comparison with \citet{GF2021}.

Although we have increased the number of DBVs, we have not found pulsations in any new objects with detected trace H. We have placed NOV limits on nine new DBAs, all of which lie beyond the theoretical red edge after 3D corrections, but their small numbers still prevent any definitive claim about whether the DBA instability strip occurs at lower temperatures than the pure-He DB instability strip. Considerable effort will be required to find more of these objects.

\section{Pulsation Properties of the DBVs} \label{sec:properties}

Pulsations in DBVs, just like their H-atmosphere DAV counterparts, are gravity modes excited by the ``convective driving'' mechanism \citep{Brickhill1991,Wu1999}. Driving is strongest for pulsation modes whose periods are about 25 times the thermal timescale at the base of the convection zone \citep{Goldreich1999}, which becomes longer as white dwarfs cool monotonically through the instability strip and the convection zone deepens. Observed pulsation periods vary between about two minutes near the hot edges of both the DA and DB instability strips, to about 20 minutes near the cool edges. The ensemble properties of DAVs, in particular how the observed pulsation modes evolve from short to long periods as a function of decreasing \teff, have been investigated for decades \citep{Robinson1979,McGraw1980,WingetFontaine1982,Clemens1993,Clemens1994,Mukadam2006}, with more recent works using {\em K2} observations of DAVs \citep{Hermes_2017_1}, homogeneous spectroscopic samples \citep{Fuchs2017}, and searches for DAVs in the {\em Gaia} survey \citep{Vincent2020} showing similar period versus \teff trends.

With the relatively small number of DBVs known prior to this work, and the large uncertainties in their effective temperatures, the ensemble pulsation properties of DBVs have yet to be investigated in great detail. While they ought to mirror those of DAVs given the similarity in driving mechanism, some complicating factors exist, such as the presence of trace atmospheric H, which can have a systematic effect on \teff. In this section, we characterize the pulsation properties of all DBVs by investigating how both the periods and amplitudes of the observed pulsation modes vary as a function of \teff. 

A common metric used to characterize the observed pulsations in white dwarfs is the weighted mean period \citep[WMP,][]{Clemens1993}, defined as $\mathrm{WMP} = \sum_{i} A_{i}P_{i} / \sum_{i} A_{i}$, where each measured pulsation period, $P_i$, is summed while being weighted by its associated amplitude, $A_i$. \citet{Mukadam2006} showed that the WMP for DAVs exhibits a roughly linear trend with \teff with a slope of $-0.83{\pm}0.08{\;}\mathrm{s\,K}^{-1}$, though more detailed studies with a larger number of homogeneously characterized DAVs \citep{Fuchs2017} show a more piece-wise linear trend with WMP first increasing more slowly between the blue edge and $T_\mathrm{eff} \simeq  12,000$\,K. Regardless of the exact trend, the WMP is a model-independent quantity that is much easier to measure with high precision compared to \teff and \logg, making it an attractive parameter with which to map properties of DBVs throughout the instability strip given their large \teff uncertainties.  
We calculate the WMP for each of the new DBVs using the periods summarized in Table~\ref{tab:periods}, and for the known DBVs using periods identified in the literature (see references in Table~\ref{tab:known_DBVs}). When calculating the WMP, we use only the independent pulsation modes, ignoring any frequencies likely to be linear combinations or harmonics. For each object, we also keep track of the period of the highest amplitude mode, $P_{\mathrm{max}}$, as a separate but similar metric which may exhibit a trend with \teff.

\begin{figure}[t!]
	\epsscale{1.1}
	\plotone{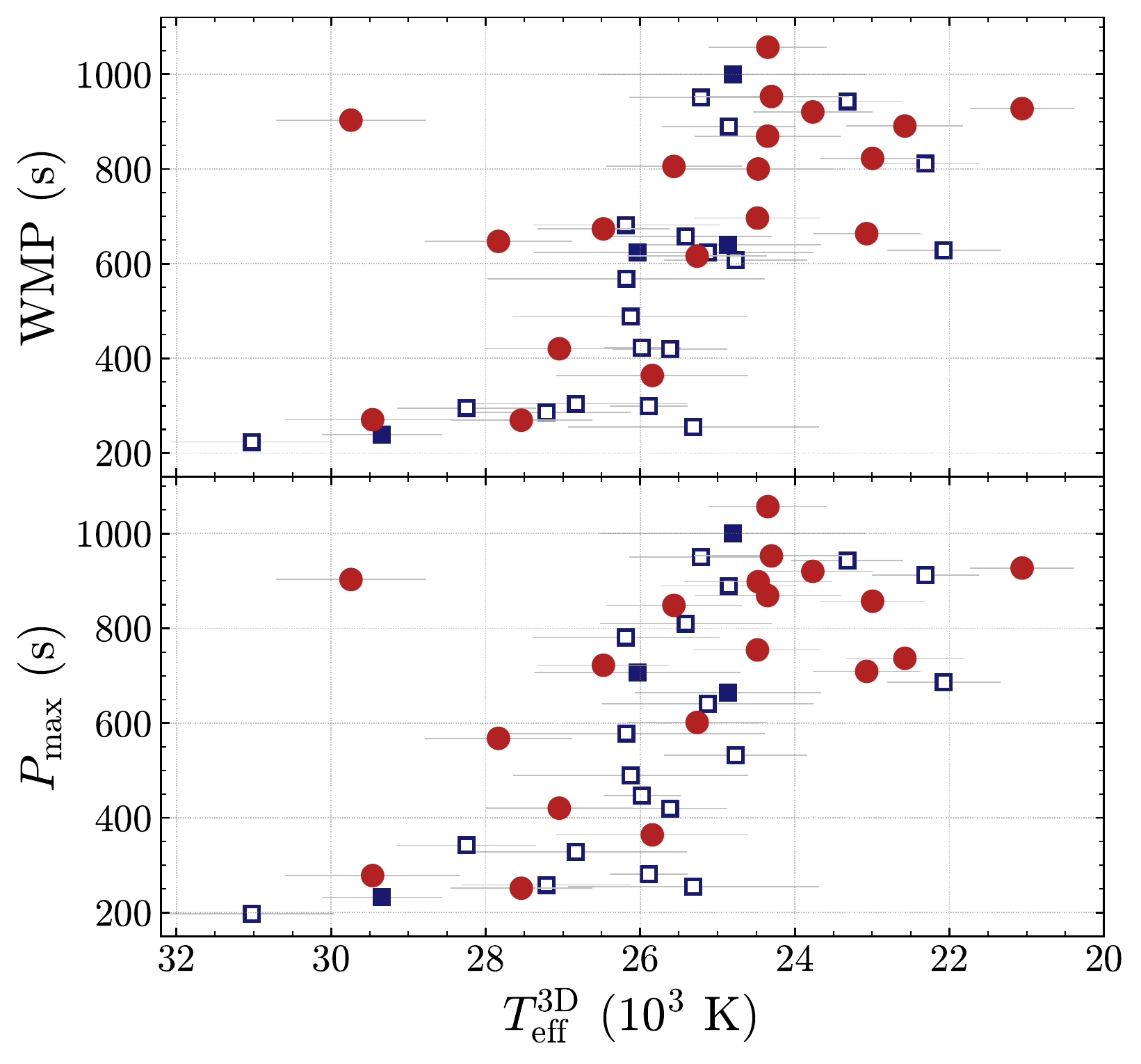}
	\caption{The weighted mean period (WMP, top panel) and highest amplitude period ($P_{\mathrm{max}}$, bottom panel) versus the 3D-corrected effective temperatures for each new (red circles) and previously known (blue squares) DBV. No harmonics or combination frequencies were used to calculate the WMP. Periods and amplitudes for previously known DBVs were taken from the literature. Both show mild trends of increasing period with decreasing effective temperature, similar to the trend observed for DAVs \citep{Robinson1979,McGraw1980,Clemens1993,Mukadam2006,Hermes_2017_1,Fuchs2017,Vincent2020}. Here and in Figure~\ref{fig:power} we adopt the \citet{Kepler_2019_1} spectroscopic solution for EPIC\,228782059 ($T_{\mathrm{eff}}\,{=}\,28{,}260\,$K), but cool solutions between 21,000 and 22,000\,K have also been proposed based on both spectroscopic and asteroseismic modeling \citep{KK2015,Duan2021}.   \label{fig:wmp}}
\end{figure}

\begin{figure}[t!]
	\epsscale{1.1}
	\plotone{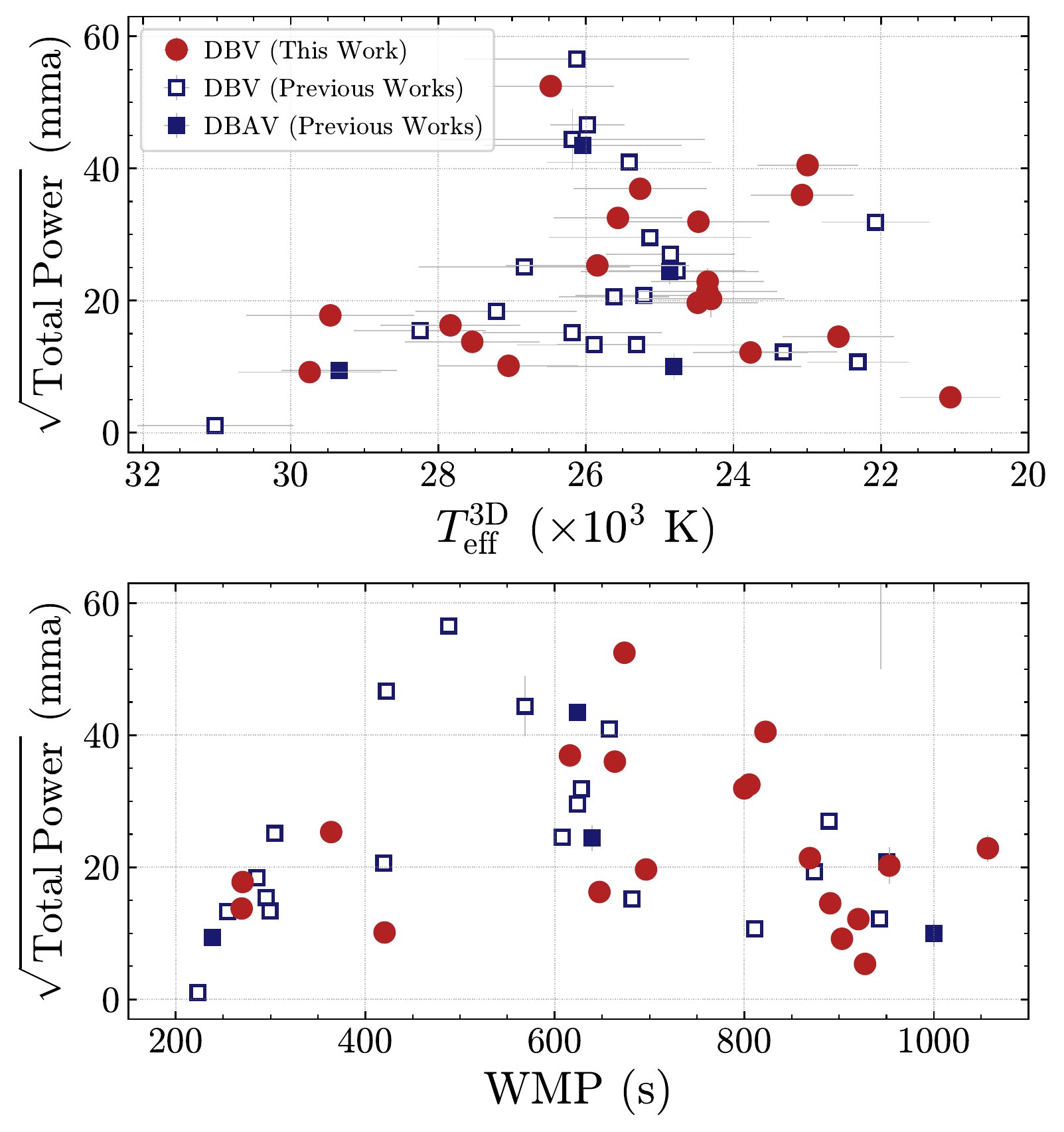}
	\caption{The square root of the maximum power mode versus \teff ({\em top panel}) and versus the weighted mean period (WMP, {\em bottom panel}) for each new and previously known DBV. Both panels appear to show a rise in pulsation power from the hot edge towards the middle of the instability strip, followed by a potential decrease towards the cool edge. These trends are again similar to those observed for DAVs \citep{Robinson1979,McGraw1980,Clemens1993,Mukadam2006,Vincent2020}. \label{fig:power}}
\end{figure}

We list the WMP and $P_{\mathrm{max}}$ values for each new DBV identified in this work in Table~\ref{tab:DBV_Params}, and in Figure~\ref{fig:wmp} we plot both the WMP and $P_{\mathrm{max}}$ versus the 3D-corrected effective temperatures. The parameters show similar trends, increasing as a function of decreasing effective temperature. Linear trends fit to both the WMP and $P_{\mathrm{max}}$ data have nearly identical slopes, with equations of $\mathrm{WMP}\,{=}\,{-}0.087^{\pm0.014}T_{\mathrm{eff}}^{\mathrm{3D}}+2840^{\pm360}$ and $P_{\mathrm{max}}\,{=}\,{-}0.091^{\pm0.015}T_{\mathrm{eff}}^{\mathrm{3D}}+2940^{\pm370}$, respectively. Compared to the slope measured by \citet{Mukadam2006} for DAVs, the WMP for DBVs exhibits a much more gradual increase with decreasing \teff. This is expected, as the thermal timescale at the base of the convection zone, $\tau_{\mathrm{th}}$, has a much weaker dependence on \teff for DBVs ($\tau_{\mathrm{th}}\propto T_{\mathrm{eff}}^{-20}$) compared to DAVs ($\tau_{\mathrm{th}}\propto T_{\mathrm{eff}}^{-90}$, \citealt{Goldreich1999,Wu2001,Montgomery2005}).

Unfortunately, using the above equations to translate a WMP or $P_{\mathrm{max}}$ value into an effective temperature is still fraught with difficulty. Even though the WMP values are model-independent, the linear trends are not since they depend on which atmospheric models, fitting procedures, and data were used to derive \teff. For example, if we redo the fitting process using only DBVs with spectroscopic parameters from \citet{Genest2019_1}, the WMP and $P_{\mathrm{max}}$ equations then become $\mathrm{WMP}\,{=}\,{-}0.108^{\pm0.021}T_{\mathrm{eff}}^{\mathrm{3D}}+3460^{\pm530}$ and $P_{\mathrm{max}}\,{=}\,{-}0.112^{\pm0.021}T_{\mathrm{eff}}^{\mathrm{3D}}+3580^{\pm550}$, respectively. The steeper trends are caused by many of the coolest DBVs in our sample having systematically higher \teff determinations in \citet{Genest2019_1}. Still, the similarities in the trends even when using atmospheric parameters from different studies suggests that the WMP, just like for DAVs, is a good proxy for \teff and can be used to investigate other trends throughout the instability strip, such as pulsation power.

As mentioned in Section~\ref{sec:DBstrip}, we find one of the new DBVs, SDSS\,J012752.18$+$140622.9, to be a significant outlier, having a relatively long WMP of $903.1$\,$\pm$\,$0.5$\,seconds, but a 3D effective temperature of $29{,}740$\,$\pm$\,$970$\,K placing it near the blue edge (the object has a photometric fit from \citealt{GF2021} using {\em Gaia} eDR3 photometry of $27{,}440$\,$\pm$\,$7900$\,K). The WMP suggests this object is much cooler, with most DBVs near this period having \teff between 22,000 and 25,000\,K. As mentioned previously, DBVs often have degenerate hot and cool solutions centered on the middle of the instability strip due to the insensitivity of He~I lines to changes in \teff in this region, so it is possible for this object to in fact be a cool DBV whose hot solution was slightly preferred during the fitting process. In \citetalias{KK2015} and \citet{Kepler_2019_1}, photometric fits to SDSS $ugriz$ were used to choose between degenerate hot and cool spectroscopic solutions. The cool solution for this object gives a 3D-corrected effective temperature of $21{,}260\,$K. Another possibility is that this object might be an unresolved double degenerate which can produce unreliable temperature determinations. The {\em Gaia} eDR3 photometry and parallax suggest this system might be over-luminous, though with low confidence given the large relative parallax uncertainty of 42\%.

Another notable object within the WMP diagram is EPIC\,228782059, a DBV that was observed during {\em K2} campaign 10 and recently proposed as possibly the coolest known DBV based on spectroscopic and asteroseismic analyses \citep{Duan2021}. In \citetalias{KK2015}, the best-fit spectroscopic model occurs at $T_{\mathrm{eff}}\,{=}\,$20,900\,K and $\log(g)\,{=}\,$7.91, while in \citet{Duan2021} the best-fit asteroseismic model occurs at $T_{\mathrm{eff}}\,{=}\,$21,900\,K and $\log(g)\,{=}\,$7.94, both consistent with EPIC\,228782059 being a cool DBV. Using the list of pulsation periods reported in \citet{Duan2021}, however, this object has a WMP of 295\,s, which as a cool DBV would make it a rather severe outlier in our WMP diagram. Similar outliers have also been found among the DAVs without a definitive explanation \citep{Mukadam2004, Fuchs2017, Vincent2020}, though an unresolved double degenerate is again one possibility. This object, however, was also included in the analyses of both \citet{Kong2018} and \citet{Kepler_2019_1} who, when fitting the same SDSS spectrum as \citetalias{KK2015}, find that hotter solutions between 28,000 and 30,000\,K are preferred. As noted by \citep{Duan2021}, these hotter solutions translate to luminosity distances that are in disagreement with the distance derived from {\em Gaia} eDR3 parallax \citep{Bailer-Jones2021}, thus favoring the cooler solutions. We adopt the hotter solution for this object from \citet{Kepler_2019_1} in Figures~\ref{fig:wmp} and \ref{fig:power}, but note that there is still considerable ambiguity about the temperature of this object.

Following the analysis of \citet{Mukadam2006} for the DAVs, we also attempt to measure the power contained within the observed pulsation modes for all of the known DBVs. Measuring the intrinsic amplitudes of pulsation modes is a much more difficult task, however, than measuring the pulsation periods, and we stress here that we are not using a homogeneous set of observations for these measurements. Pulsation amplitudes are wavelength dependent, so observations in different filters will affect the measured amplitudes. Also, in our relatively short McDonald runs, any closely spaced modes from successive radial overtones or rotational splittings will remain unresolved and beat with one another, producing amplitudes that are higher or lower than if they were resolved. Periods and amplitudes for several previously known DBVs, however, come from extensive observations using the Whole Earth Telescope, the {\em Kepler} spacecraft, and the {\em Transiting Exoplanet Survey Satellite} ({\em TESS}), which are able to resolve closely spaced modes that single-night McDonald runs cannot.

Even in the absence of observational limitations, the intrinsic amplitudes remain difficult to determine due to geometric cancellation caused by disk averaging, the inclination angle of the white dwarf, and limb darkening. Pulsations in white dwarfs produce temperature variations on the surface that can be described using spherical harmonics \citep{RKN1982}, with indices $\ell$ and $m$ describing the number and distribution of pulsation nodes across the surface of the star. Modes with higher $\ell$ exhibit a larger number of bright and dark regions that, when averaged over the disk of the white dwarf, experience more cancellation. The inclination of the white dwarf determines the distribution of bright and dark spots that fall within our field of view for a given mode, while limb darkening reduces the amount of flux coming from the edges of the stellar disk. Disk averaging and the inclination angle will always serve to reduce the measured amplitude with respect to the intrinsic amplitude, though limb darkening actually boosts the measured amplitude by reducing the amount of cancellation between bright and dark spots seen near the limb of the white dwarf.

All of these factors, in addition to the highly uncertain temperatures for DBVs, make the interpretation of a pulsation power versus \teff diagram difficult. Still, we present a current best-effort attempt at producing such a diagram for DBVs so that we can compare with their DAV counterparts. We calculate the total pulsation power ($p$) for each object by summing the power of each independent mode, $p = \sum_{i}A_i^{2}$, again excluding any linear combinations or harmonics. In Figure~\ref{fig:power} we present the square root of the total power versus \teff in the top panel, and versus the WMP in the bottom panel.

In both panels of Figure~\ref{fig:power}, the total pulsation power appears to increase from the blue edge to the middle of the instability strip, and then fall back down again at the red edge. This is similar to the DAVs, whose rise in pulsation power from the blue edge has been documented for decades \citep{Robinson1979,Robinson1980,McGraw1980,Fontaine1982,Clemens1993,Kanaan2002}, while the decrease in power close to the red edge was only observed more recently \citep{Mukadam2006,Vincent2020}. In combination with the WMP versus \teff trends, these qualitative similarities support the idea that DAV and DBV pulsations are being driven by similar mechanisms. Perhaps the main difference seen in the DBV pulsation power, however, is that the peak might happen at lower WMP before falling off. In the DAVs, pulsation power begins decreasing between WMPs of 900 and 1000 seconds, while for DBVs it appears to happen somewhat sooner, around 800 seconds. This small difference, however, might just be a matter of still having too few DBVs to properly sample the power versus WMP diagram.

\section{Conclusions} \label{sec:conclude}

We obtained time series photometry for 55 DB and DBA white dwarfs located in and near the DB instability strip based on atmospheric parameters determined from SDSS spectra. Of these, we found 19 DBs to pulsate, and placed limits on variability, often less than 0.5\%, for the remaining 36 objects. Compared to the 28 previously known DBVs, the new DBVs presented here do not significantly extend the DB instability strip in either the hot or cool directions, but improve constraints on the empirical locations of the red (cool) and blue (hot) edges, especially given how uncertain the temperatures for these objects typically are.

After applying the 3D convection corrections determined by \citet{Cukanovaite2021} to spectroscopic \teff and \logg from the literature, we find that the most recent theoretical calculations describing the blue and red edge locations \citep{vanGrootel_2017_1} agree well with the empirical DB instability strip, although the observed blue and red edges both appear hotter than the respective theoretical edges. Even so, we caution that we have not used a homogeneous spectroscopic sample, and the differences in atmospheric models, fitting procedures, and observational data quality between objects can influence the location of the instability strip. In the second paper of this series, we plan to present a homogeneous spectroscopic study of numerous DBVs and NOVs using observations from the Hobby Eberly Telescope at McDonald Observatory.

We found several NOVs within the theoretical instability strip which can be accounted for with a variety of explanations. Given their large temperature uncertainties, some NOVs may in fact lie outside the instability strip. For others, perhaps the pulsations were undergoing destructive beating during our observations, suppressing the observed pulsation amplitudes below our detection thresholds. Lastly, some DBVs might just have low-amplitude pulsations, which is especially common among the blue-edge pulsators like PG\,0112+104. More extensive observations would be required in most cases to rule out pulsations more definitively before assessing the purity of the DB instability strip.

With the larger number of DBVs now available, we presented the first analysis of the ensemble properties of DBVs, investigating how the weighted mean period and total pulsation power change as a function of effective temperature. We find both to exhibit qualitatively similar trends when compared with the DAVs \citep{Clemens1993,Mukadam2006}, with the weighted mean period increasing gradually with decreasing \teff, and the pulsation power initially increasing towards the middle of the instability strip before potentially decreasing towards the cool edge. The similarities in pulsation properties between DAVs and DBVs supports the idea that they have similar driving mechanism.

To further improve constraints on the observed blue and red edges, increasing the number of DBVs is still important, as is performing homogeneous spectroscopic or photometric analyses of as many DBVs and NOVs as possible for \teff, \logg, and [H/He] determinations. After our search, only 40 relatively bright DBs with $g<19$~mag and \teff between 22,000 and 29,000\,K remain in the \citetalias{KK2015} catalogue that have not yet been assessed for variability. Future spectroscopic surveys covering a much larger area of the sky will be vital for future DBV searches as they will increase the number of spectroscopically confirmed DBs near to and within the instability strip. For example, the SDSS-V Milky Way Mapper plans to obtain hundreds of thousands of optical spectra of white dwarfs as part of its ``White Dwarf Chronicle'' survey \citep{Kollmeier2017}, an order-of-magnitude increase in the number of spectroscopically observed white dwarfs. Using variability metrics based on Zwicky Transient Facility and {\em Gaia} photometry \citep[e.g.,][]{Guidry2021} may also provide a promising method to efficiently identify some high amplitude DBVs without prior spectral classifications.

\begin{acknowledgments}

We thank the referee for their insightful comments that helped improve this work. Z.P.V., D.E.W., and M.H.M. acknowledge support from the United States Department of Energy under grant DE-SC0010623, the National Science Foundation under grant AST-1707419, and the Wootton Center for Astrophysical Plasma Properties under the United States Department of Energy collaborative agreement DE-NA0003843. Z.P.V. also acknowledges the Heising-Simons Foundation for postdoctoral scholar support at Caltech, and M.H.M. acknowledges support from the NASA ADAP program under grant 80NSSC20K0455.
K.J.B. is supported by the National Science Foundation under Award AST-1903828. Data from McDonald Observatory were obtained with financial support from NASA K2 Cycle 5 Grant 80NSSC18K0387 and WCAPP. S.O.K. acknowledges support from Coordena\c{c}\~ao de Aperfei\c{c}oamento de Pessoal de N\'{\i}vel Superior - Brasil (CAPES) - Finance Code 001, Conselho Nacional de Desenvolvimento Cient\'{\i}fico e Tecnol\'ogico - Brasil (CNPq), and Funda\c{c}\~ao de Amparo \`a Pesquisa do Rio Grande do Sul (FAPERGS) - Brasil. We thank McDonald Observatory and all of the observing support staff who helped make these observations possible.
 
\software{Astropy \citep{Astropy_2018},
          {\sc iraf} (National Optical Astronomy Observatories),
          {\sc Period04} \citep{Period04},
          {\sc Wqed} \citep{Thompson2013}, 
          {\tt ccd\_hsp} \citep{Kanaan2002},
          and the NASA Astrophysics Data System (ADS) repositories.
          }
          
\end{acknowledgments}
 
\bibliography{ref}
\bibliographystyle{aasjournal}

\appendix 

\section{Additional Tables} \label{appendix}

Table~\ref{tab:obs} provides a summary of time series photometric observations from McDonald Observatory. Table~\ref{tab:DBV_Params} provides the atmospheric parameters and some pulsation properties for the DBVs. Table~\ref{tab:NOV_Params} provides the atmospheric properties and variability limits for the NOVs.

\begin{deluxetable*}{lccc|lccc|lccc}[b]
\tablecaption{Summary of McDonald 2.1-m Observations \label{tab:obs}}
\tabletypesize{\scriptsize}
\tablecolumns{12}
\tablewidth{0pt}
\tablehead{
\colhead{Name} & \colhead{Date} & \colhead{$\Delta$T} & \colhead{$t_\mathrm{exp}$}  &
\colhead{Name} & \colhead{Date} & \colhead{$\Delta$T} & \colhead{$t_\mathrm{exp}$}  &
\colhead{Name} & \colhead{Date} & \colhead{$\Delta$T} & \colhead{$t_\mathrm{exp}$}  \\ [-0.2cm]
\colhead{(SDSS J)} & \colhead{} & \colhead{(hr)} & \colhead{(s)} &
\colhead{(SDSS J)} & \colhead{} & \colhead{(hr)} & \colhead{(s)} &
\colhead{(SDSS J)} & \colhead{} & \colhead{(hr)} & \colhead{(s)}
}
\startdata
002458.42{\tt+}245834.2       & 2017 Jul 28 & 2.13 & 10 & 082316.32{\tt+}233317.8       & 2017 Dec 12 & 1.68 & 10 &                               & 2018 May 16 & 2.82 &  5  \\
                              & 2017 Jul 30 & 1.69 & 10 &                               & 2017 Dec 16 & 1.79 & 20 & 154201.50{\tt+}502532.1       & 2018 May 12 & 1.15 &  5  \\
                              & 2017 Oct 21 & 4.16 & 20 & 083035.14{\tt+}564459.4{\bf*} & 2017 Dec 12 & 1.31 & 15 &                               & 2018 May 13 & 2.92 &  5  \\
011607.92{\tt+}330154.3{\bf*} & 2017 Dec 12 & 1.80 & 15 &                               & 2018 Jan 23 & 1.07 & 15 & 155327.56{\tt+}150545.7{\bf*} & 2018 May 13 & 2.86 & 15  \\
                              & 2017 Dec 13 & 1.86 & 15 & 083415.45{\tt+}254819.9{\bf*} & 2018 Mar 14 & 1.11 & 15 & 155921.08{\tt+}190407.8       & 2017 May 19 & 2.84 & 10  \\
012752.18{\tt+}140622.9{\bf*} & 2018 Sep 10 & 3.45 & 10 &                               & 2018 Mar 16 & 3.11 & 15 &                               & 2017 May 21 & 0.77 &  5  \\
                              & 2017 Sep 11 & 3.92 & 10 & 084211.30{\tt+}461819.0{\bf*} & 2017 Dec 14 & 2.76 & 15 &                               & 2017 May 22 & 2.22 & 15  \\
014945.65{\tt+}223016.4       & 2018 Sep 12 & 1.79 & 15 & 084350.85{\tt+}361419.5       & 2017 Oct 24 & 2.48 & 30 & 162425.01{\tt+}295511.8{\bf*} & 2018 Jun 23 & 2.03 & 15  \\
                              & 2018 Sep 13 & 1.96 & 15 &                               & 2017 Oct 25 & 2.00 & 10 & 165349.37{\tt+}274647.3{\bf*} & 2018 Aug 14 & 1.61 & 15  \\
                              & 2018 Sep 14 & 0.71 & 10 & 084614.89{\tt+}193515.3       & 2017 Dec 13 & 1.98 & 15 &                               & 2018 Aug 15 & 3.34 & 15  \\
020409.84{\tt+}212948.5       & 2017 Jul 29 & 1.68 & 10 & 084953.09{\tt+}105621.2       & 2019 May 08 & 2.09 & 10 &                               & 2018 Aug 16 & 3.10 & 15  \\
                              & 2017 Jul 30 & 2.47 & 15 &                               & 2019 May 09 & 2.26 & 10 &                               & 2018 Oct 10 & 2.15 & 15  \\
023402.50{\tt+}243352.2       & 2017 Oct 23 & 2.29 & 20 & 092106.44{\tt+}140736.7       & 2017 Dec 13 & 2.01 & 10 &                               & 2018 Oct 11 & 2.21 & 15  \\
                              & 2017 Oct 26 & 1.38 & 30 &                               & 2018 May 14 & 2.77 & 15 & 173232.09{\tt+}335610.4{\bf*} & 2018 Jul 12 & 2.47 & 15  \\
025352.96{\tt+}332803.6{\bf*} & 2017 Dec 12 & 2.08 & 15 &                               & 2018 May 16 & 1.91 & 15 &                               & 2018 Jul 17 & 1.12 & 15  \\
                              & 2017 Dec 13 & 3.64 & 15 & 092355.26{\tt+}085717.3       & 2019 Dec 15 & 2.54 & 20 & 174025.00{\tt+}245705.5       & 2018 Jun 23 & 2.73 & 15  \\
065146.31{\tt+}271927.3       & 2017 Nov 22 & 2.12 & 15 &                               & 2019 Dec 16 & 3.52 & 15 &                               & 2018 Jun 24 & 4.15 & 15  \\
                              & 2017 Nov 23 & 1.46 & 10 & 101502.95{\tt+}464835.3{\bf*} & 2018 Jan 24 & 2.89 & 10 & 183252.20{\tt+}421526.1       & 2018 Jun 19 & 2.93 & 15  \\
073935.14{\tt+}244505.2{\bf*} & 2017 Oct 20 & 1.07 & 15 & 105423.94{\tt+}211057.4       & 2018 Jan 23 & 3.17 & 10 &                               & 2018 Jun 25 & 2.98 & 15  \\
                              & 2017 Oct 21 & 1.47 & 10 &                               & 2018 Jan 28 & 2.52 & 10 & 212403.12{\tt+}114230.2{\bf*} & 2018 Jun 20 & 2.79 & 15  \\
074925.14{\tt+}195040.0       & 2017 Nov 24 & 1.74 & 10 & 110235.85{\tt+}623416.1{\bf*} & 2018 Mar 17 & 3.37 &  8 &                               & 2018 Jun 21 & 3.10 & 15  \\
                              & 2017 Dec 13 & 1.46 & 15 &                               & 2020 Jan 13 & 4.09 & 20 &                               & 2018 Jun 22 & 3.31 & 10  \\
075452.85{\tt+}194907.0       & 2017 Dec 17 & 4.12 & 15 & 112752.92{\tt+}553522.0       & 2017 May 21 & 3.13 & 15 & 214441.71{\tt+}010029.8       & 2017 Jul 31 & 2.52 & 15  \\
                              & 2018 Jan 24 & 1.80 &  5 &                               & 2017 May 23 & 1.67 & 15 &                               & 2018 Aug 16 & 2.39 & 15  \\
075523.86{\tt+}172825.1       & 2017 Oct 21 & 1.24 & 15 & 113247.25{\tt+}283519.0       & 2018 Mar 13 & 1.15 &  5 & 220250.26{\tt+}213120.2       & 2017 May 23 & 1.08 & 15  \\
                              & 2017 Oct 22 & 1.96 & 15 &                               & 2018 Mar 16 & 2.64 & 15 &                               & 2017 May 24 & 1.67 & 20  \\
080236.92{\tt+}154813.6{\bf*} & 2017 Oct 26 & 1.91 & 20 & 131646.02{\tt+}414639.0       & 2018 Mar 14 & 1.36 & 15 & 222833.82{\tt+}141036.9       & 2017 Nov 23 & 2.74 & 15  \\
                              & 2018 Jan 23 & 2.09 & 10 &                               & 2018 May 16 & 2.80 &  5 &                               & 2017 Nov 24 & 2.07 & 10  \\
080349.15{\tt+}085532.6       & 2017 Nov 24 & 1.38 & 10 & 140028.43{\tt+}475644.1       & 2017 Jun 24 & 2.19 & 30 & 225020.91{\tt-}091425.6{\bf*} & 2018 Aug 15 & 4.82 & 15  \\
                              & 2017 Dec 12 & 1.40 & 15 &                               & 2017 Jun 25 & 2.34 & 15 & 225424.73{\tt+}231515.8{\bf*} & 2018 Jul 19 & 1.30 & 15  \\
081345.42{\tt+}365140.5{\bf*} & 2017 Dec 18 & 4.23 & 15 & 142405.54{\tt+}181807.3       & 2019 May 01 & 2.62 & 15 &                               & 2018 Jul 20 & 4.92 & 15  \\
                              & 2017 Dec 21 & 3.27 & 15 &                               & 2019 May 02 & 3.91 & 15 & 232108.40{\tt+}010433.5       & 2017 Nov 25 & 1.85 & 15  \\
                              & 2018 Jan 25 & 2.92 & 10 & 144814.33{\tt+}150449.7       & 2017 Jun 26 & 1.32 & 30 &                               & 2017 Dec 16 & 2.22 & 10  \\
081453.55{\tt+}300734.8{\bf*} & 2017 Nov 24 & 1.55 & 14 &                               & 2017 Jun 27 & 3.14 & 10 & 232711.11{\tt+}515344.7       & 2018 Sep 10 & 4.00 &  3  \\
                              & 2020 Jan 23 & 2.72 & 30 & 145755.43{\tt+}015442.9       & 2017 May 19 & 1.78 & 10 &                               & 2018 Sep 11 & 4.00 &  3  \\
081656.17{\tt+}204946.0       & 2017 Nov 22 & 1.92 & 20 &                               & 2017 May 23 & 2.66 & 15 &                               & 2018 Sep 12 & 3.01 &  3  \\
                              & 2017 Nov 25 & 1.78 & 20 & 151729.46{\tt+}433028.6       & 2018 Jul 06 & 1.55 & 15 & 234848.77{\tt+}381754.6       & 2017 Nov 23 & 1.86 &  5  \\
                              & 2017 Dec 12 & 1.28 & 10 & 153454.99{\tt+}224918.6       & 2018 May 14 & 2.51 & 15 &                               & 2017 Nov 24 & 1.03 &  3  \\
\enddata
\tablenotetext{}{{\bf*}New DBVs from this work}
\end{deluxetable*}

\begin{deluxetable*}{lccccccc|cc}
\tablecaption{Atmospheric Parameters and Pulsation Properties for New DBVs \label{tab:DBV_Params}}
\tablecolumns{01}
\tabletypesize{\scriptsize}
\tablehead{
\colhead{Name} & 
\colhead{Type} &
\colhead{P-M-F} & 
\colhead{$g$} &
\colhead{$T_{\mathrm{eff}}$} & 
\colhead{$\log(g)$} & 
\colhead{$T_{\mathrm{eff}}^{3\mathrm{D}}$} & 
\colhead{$\log(g)^{3\mathrm{D}}$} &
\colhead{WMP} &
\colhead{$P_{\mathrm{max}}$} \\ [-0.2cm]
\colhead{(SDSS J)} & \colhead{} &
\colhead{} & \colhead{(mag)} & 
\colhead{(K)} &
\colhead{(cgs)} &
\colhead{(K)} & 
\colhead{(cgs)} &
\colhead{(s)} & 
\colhead{(s)}
}
\startdata
 011607.92+330154.3 & DBV & 6594-56272-971 & 18.9 & 21040[760]  & 8.03[0.13] & 22580 & 7.98 & 890.7[0.6] & 736.8[0.6] \\
 012752.18+140622.9 & DBV & 4665-56209-726 & 18.3 & 29760[970]  & 7.99[0.13] & 29740 & 7.99 & 903.1[0.5] & 903.1[0.7] \\
 025352.96+332803.6 & DBV & 2398-53768-185 & 18.8 & 27560[920]  & 7.72[0.13] & 27540 & 7.73 & 269.6[0.2] & 251.4[0.2] \\
 073935.14+244505.2 & DBV & 4470-55587-626 & 17.3 & 21500[700]  & 7.89[0.12] & 23070 & 7.86 & 663.4[0.3] & 709.2[0.3] \\
 080236.92+154813.6 & DBV & 4494-55569-174 & 17.4 & 21400[690]  & 7.97[0.12] & 22990 & 7.93 & 822.3[0.9] & 857.4[1.0] \\
 081345.42+365140.5 & DBV & 2674-54097-287 & 18.8 & 27060[950]  & 7.61[0.13] & 27050 & 7.61 & 420.4[0.9] & 420.4[0.8] \\
 081453.55+300734.8 & DBV & 930-52618-565  & 18.7 & 22630[940]  & 8.02[0.13] & 24350 & 8.00 & 869.2[2.5] & 869.2[2.5] \\
 083035.14+564459.4 & DBV & 1783-53386-540 & 17.3 & 26490[850]  & 7.82[0.12] & 26480 & 7.81 & 673.5[1.0] & 722.2[1.1] \\
 083415.45+254819.9 & DBV & 1930-53347-357 & 18.3 & 22850[960]  & 7.98[0.13] & 24470 & 7.96 & 799.9[1.1] & 898.5[1.3] \\
 084211.30+461819.0 & DBV & 763-52235-435  & 18.5 & 24780[880]  & 8.00[0.13] & 25570 & 7.99 & 805.3[1.2] & 848.6[1.3] \\
 101502.95+464835.3 & DBV & 944-52614-328  & 18.6 & 23460[810]  & 7.82[0.13] & 24480 & 7.80 & 696.3[1.2] & 754.4[1.6] \\
 110235.85+623416.1 & DBV & 2882-54498-20  & 17.7 & 23160[760]  & 7.83[0.12] & 24350 & 7.81 & 1057.0[1.9] & 1057.0[1.9] \\
 155327.56+150545.7 & DBV & 2521-54538-276 & 18.1 & 24720[900]  & 7.90[0.13] & 25260 & 7.89 & 616.0[0.8] & 601.3[0.9] \\
 162425.01+295511.8 & DBV & 4953-55749-422 & 18.0 & 22430[780]  & 7.78[0.13] & 23770 & 7.75 & 920.4[2.5] & 920.4[2.5] \\
 165349.37+274647.3 & DBV & 1690-53475-637 & 18.7 & 19800[680]  & 7.67[0.14] & 21060 & 7.61 & 927.4[0.6] & 927.4[0.6] \\
 173232.09+335610.4 & DBV & 2262-54623-450 & 19.3 & 23010[1000] & 7.84[0.14] & 24300 & 7.82 & 953.0[4.1] & 953.0[4.1] \\
 212403.12+114230.2 & DBV & 730-52466-380  & 19.0 & 29480[1140] & 7.75[0.15] & 29460 & 7.75 & 270.6[0.1] & 278.1[0.1] \\
 225020.91-091425.6 & DBV & 724-52254-341  & 18.8 & 25390[1240] & 8.01[0.14] & 25840 & 8.01 & 364.0[0.4] & 364.0[0.4] \\
 225424.73+231515.8 & DBV & 6308-56215-843 & 19.0 & 27850[950]  & 7.85[0.13] & 27830 & 7.86 & 647.1[0.3] & 567.8[0.3] \\
\enddata
\tablecomments{All new DBVs have only upper limits on [H/He] from \citetalias{KK2015} or \citet{Kepler_2019_1}. 3D corrections were calculated assuming [H/He]\,=\,$-$10.}
\end{deluxetable*}

\begin{deluxetable*}{llccccccc|cc}
\tablecaption{Atmospheric Parameters and Variability Limits for New NOVs \label{tab:NOV_Params}}
\tablecolumns{11}
\tablewidth{0pt}
\tabletypesize{\scriptsize}
\tablehead{
\colhead{Name} & 
\colhead{Type} &
\colhead{P-M-F} & 
\colhead{$g$} &
\colhead{$T_{\mathrm{eff}}$} & 
\colhead{$\log(g)$} & 
\colhead{[H/He]} &
\colhead{$T_{\mathrm{eff}}^{3\mathrm{D}}$} & 
\colhead{$\log(g)^{3\mathrm{D}}$} &
\colhead{NOV$^{\dagger}$} & 
\colhead{Runs$^{\dagger}$} \\ [-0.2cm]
\colhead{(SDSS J)} & \colhead{} &
\colhead{} & \colhead{(mag)} & 
\colhead{(K)} &
\colhead{(cgs)} &
\colhead{} &
\colhead{(K)} & 
\colhead{(cgs)} &
\colhead{(mma)} &
\colhead{}
}
\startdata
 002458.42+245834.2 & DB  & 6279-56243-16  & 17.3 & 20690[660] & 8.07[0.12]  & ---         & 22130 & 8.01 &  2.3 & 3 \\
 014945.65+223016.4 & DB  & 5112-55895-304 & 19.3 & 31490[1120] & 7.67[0.14] & ---         & 31480 & 7.68 &  7.1 & 3 \\
 020409.84+212948.5 & DBA & 5113-55924-232 & 18.1 & 20980[660] & 8.25[0.12]  & -2.99[0.13] & 22260 & 8.23 &  3.1 & 2 \\
 023402.50+243352.2 & DBA & 2399-53764-105 & 19.0 & 29760[1070] & 7.86[0.13] & -2.76[0.08] & 29740 & 7.87 &  6.1 & 2 \\
 065146.31+271927.3 & DB  & 2694-54199-23  & 18.2 & 35800[1190] & 7.96[0.13] & ---         & 35800 & 7.96 &  4.1 & 2 \\
 074925.14+195040.0 & DB  & 4485-55836-185 & 17.9 & 19720[620] & 7.97[0.12]  & ---         & 20920 & 7.90 &  3.5 & 2 \\
 075452.85+194907.0 & DB  & 4482-55617-368 & 18.3 & 21300[710] & 7.90[0.12]  & ---         & 22850 & 7.86 &  3.3 & 2 \\
 075523.86+172825.1$^{\ddagger}$ & DB  & 2729-54419-106 & 17.9 & 27180[870] & 7.77[0.12]  & ---         & 27160 & 7.77 &  4.0 & 2 \\
 080349.15+085532.6 & DB  & 2419-54139-98  & 18.0 & 21860[760] & 7.98[0.13]  & ---         & 23520 & 7.95 &  4.3 & 2 \\
 081656.17+204946.0$^{\ddagger}$ & DB  & 1925-53327-573 & 17.0 & 27460[870] & 7.89[0.12]  & ---         & 27450 & 7.89 &  2.1 & 3 \\
 082316.32+233317.8 & DBA & 4468-55894-114 & 17.6 & 19860[630] & 8.08[0.12]  & -4.13[0.22] & 20930 & 8.05 &  3.1 & 2 \\
 084350.85+361419.5 & DB  & 4609-56251-422 & 17.0 & 20430[660] & 8.03[0.12]  & ---         & 21810 & 7.96 &  1.9 & 2 \\
 084614.89+193515.3 & DBA & 2280-53680-248 & 18.1 & 20560[670] & 8.25[0.13]  & -3.22[0.24] & 21730 & 8.23 &  5.4 & 1 \\
 084953.09+105621.2 & DB  & 2671-54141-476 & 17.9 & 30530[970] & 8.04[0.12]  & ---         & 30510 & 8.05 &  3.3 & 2 \\
 092106.44+140736.7 & DBA & 5305-55984-134 & 18.0 & 22360[730] & 8.51[0.12]  & -3.39[0.25] & 24000 & 8.49 &  2.7 & 3 \\
 092355.26+085717.3 & DBA & 1302-52763-489 & 16.4 & 20500[650] & 7.96[0.12]  & -4.16[0.15] & 21750 & 7.94 &  1.4 & 2 \\
 105423.94+211057.4 & DB  & 6427-56328-162 & 17.3 & 22010[740] & 7.95[0.12]  & ---         & 23660 & 7.92 &  1.8 & 2 \\
 112752.92+553522.0 & DBA & 7093-56657-80  & 17.1 & 19640[620] & 8.19[0.12]  & -4.17[0.13] & 20620 & 8.15 &  2.3 & 2 \\
 113247.25+283519.0$^{\ddagger}$ & DB  & 2217-53794-50  & 18.7 & 25260[990] & 7.97[0.13]  & ---         & 25680 & 7.96 &  4.1 & 2 \\
 131646.02+414639.0 & DB  & 1460-53138-535 & 17.2 & 21840[720] & 7.95[0.12]  & ---         & 23490 & 7.92 &  2.2 & 2 \\
 140028.43+475644.1 & DB  & 6750-56367-648 & 17.1 & 30830[980] & 7.92[0.12]  & ---         & 30810 & 7.92 &  2.3 & 2 \\
 142405.54+181807.3 & DB  & 2760-54506-374 & 18.8 & 30010[1100] & 8.00[0.12] & ---         & 29990 & 8.00 &  4.4 & 2 \\
 144814.33+150449.7 & DB  & 2750-54242-334 & 15.7 & 20330[640] & 7.93[0.12]  & ---         & 21690 & 7.87 &  1.9 & 2 \\
 145755.43+015442.9 & DB  & 4015-55624-316 & 18.1 & 19600[620] & 8.00[0.12]  & ---         & 20760 & 7.93 &  3.6 & 2 \\
 151729.46+433028.6 & DB  & 1678-53433-372 & 18.3 & 28930[970] & 8.00[0.12]  & ---         & 28920 & 8.00 & 11.4 & 1 \\
 153454.99+224918.6 & DB  & 2162-54207-528 & 17.6 & 20800[680] & 8.01[0.13]  & ---         & 22270 & 7.95 &  3.4 & 2 \\
 154201.50+502532.1 & DB  & 796-52401-180  & 16.8 & 31030[990] & 7.64[0.12]  & ---         & 31010 & 7.65 &  1.4 & 2 \\
 155921.08+190407.8 & DBA & 3930-55332-259 & 16.4 & 20680[650] & 7.99[0.12]  & -3.80[0.16] & 21960 & 7.97 &  2.0 & 3 \\
 174025.00+245705.5 & DBA & 2183-53536-303 & 17.5 & 20440[650] & 8.33[0.13]  & -3.31[0.19] & 21560 & 8.30 &  2.4 & 2 \\
 183252.20+421526.1 & DB  & 2819-54617-322 & 17.8 & 21400[710] & 7.92[0.12]  & ---         & 22970 & 7.87 &  2.2 & 2 \\
 214441.71+010029.8 & DB  & 4196-55478-714 & 18.3 & 20080[650] & 7.96[0.12]  & ---         & 21380 & 7.89 &  4.9 & 2 \\
 220250.26+213120.2 & DB  & 5948-56107-107 & 16.9 & 20290[640] & 7.99[0.12]  & ---         & 21640 & 7.92 &  2.2 & 2 \\
 222833.82+141036.9 & DB  & 737-52518-602  & 18.8 & 31720[1220] & 7.76[0.15] & ---         & 31710 & 7.76 &  3.6 & 2 \\
 232108.40+010433.5 & DB  & 382-51816-614  & 18.1 & 21040[870] & 8.14[0.14]  & ---         & 22570 & 8.08 &  4.1 & 2 \\
 232711.11+515344.7 & DB  & 1662-52970-96  & 17.3 & 30510[990] & 7.84[0.13]  & ---         & 30490 & 7.84 &  1.3 & 3 \\
 234848.77+381754.6$^{\ddagger}$ & DB  & 1882-53262-136 & 17.5 & 23920[880] & 8.06[0.13]  & ---         & 25400 & 8.05 &  2.1 & 2 \\
\enddata
\tablecomments{SDSS\,J065146.31+271927.3 falls outside the 3D correction grids of \citet{Cukanovaite2021}, so the 3D corrected values for this object are the same as the 1D values. The spectral types displayed are based solely on the detection (DBA) or non-detection (DB) of trace H in the SDSS spectra analyzed by \citetalias{KK2015} and \citet{Kepler_2019_1}, and may differ from previous spectral classifications.}
\vspace{-0.2cm}
\tablenotetext{\dagger}{NOV represents the not-observed-to-vary limit from our McDonald runs, while the number of runs indicates how many individual nights an object was observed not to vary.}
\vspace{-0.2cm}
\tablenotetext{\ddagger}{NOVs more than 1$\sigma$ inside the theoretical instability strip and with variability limits ${<}5\,$mma, as shown in Figure~\ref{fig:ampstrip}.}
\end{deluxetable*}
	
\end{document}